\documentclass{sigchi}

\usepackage{balance}       %
\usepackage{graphics}      %
\usepackage[T1]{fontenc}   %
\usepackage{txfonts}
\usepackage{mathptmx}
\usepackage[pdflang={en-US},pdftex]{hyperref}
\usepackage{color}
\usepackage{booktabs}
\usepackage{textcomp}
\usepackage{tabu}                      %
\usepackage{amsmath}

\DeclareMathOperator*{\argmin}{arg\,min}

\usepackage{color}
\usepackage{amsmath}
\usepackage{amssymb}
\usepackage{amsfonts}
\usepackage{subfigure}
\usepackage{xspace}
\usepackage{enumitem}
\usepackage{mathtools}
\usepackage{microtype}

\newcommand{\hl}[1]{\textcolor{blue}{#1}}
\renewcommand{\hl}[1]{#1}

\newcommand{\hidecomment}[1]{}

\newcommand{\f}{f}

\newcommand{\vtxhead}{v_{\text{head}}}
\newcommand{\vtxtail}{v_{\text{tail}}}

\newcommand{\old}[1]{}
\newcommand{\oursystem}{Unwind\xspace}

\AtBeginDocument{%
  \providecommand\BibTeX{{%
    \normalfont B\kern-0.5em{\scshape i\kern-0.25em b}\kern-0.8em\TeX}}}

\usepackage{microtype}        %
\usepackage{ccicons}          %

\def\plaintitle{Unwind: Interactive Fish Straightening}
\def\plainauthor{Francis~Williams, Alexander~Bock, Harish~Doraiswamy, Cassandra~Donatelli, Kayla~Hall, Adam~Summers, Daniele~Panozzo, Claudio~T.~Silva}
\def\plainkeywords{CT Scan Data, Volumetric Deformation, Interactive System}

\makeatletter
\def\url@leostyle{%
  \@ifundefined{selectfont}{
    \def\UrlFont{\sf}
  }{
    \def\UrlFont{\small\bf\ttfamily}
  }}
\makeatother
\def\pprw{8.5in}
\def\pprh{11in}

\definecolor{linkColor}{RGB}{6,125,233}
\hypersetup{%
  pdftitle={\plaintitle},
 pdfauthor={\plainauthor},
  pdfkeywords={\plainkeywords},
  pdfdisplaydoctitle=true, %
  bookmarksnumbered,
  pdfstartview={FitH},
  colorlinks,
  citecolor=black,
  filecolor=black,
  linkcolor=black,
  urlcolor=linkColor,
  breaklinks=true,
  hypertexnames=false
}

\toappear{\scriptsize Permission to make digital or hard copies of all or part of this work for personal or classroom use is granted without fee provided that copies are not made or distributed for profit or commercial advantage and that copies bear this notice and the full citation on the first page. Copyrights for components of this work owned by others than the author(s) must be honored. Abstracting with credit is permitted. To copy otherwise, or republish, to post on servers or to redistribute to lists, requires prior specific permission and/or a fee. Request permissions from permissions@acm.org. \\
{\emph{CHI '20, April 25--30, 2020, Honolulu, HI, USA.} } \\
Copyright is held by the owner/author(s). Publication rights licensed to ACM. \\
ACM ISBN 978-1-4503-6708-0/20/04\ ...\$15.00.\\
http://dx.doi.org/10.1145/3313831.3376846}

\begin{document}

\title{Unwind: Interactive Fish Straightening}

\numberofauthors{1}
\author{%
  \alignauthor{Francis~Williams\textsuperscript{1}, Alexander~Bock\textsuperscript{2}, Harish~Doraiswamy\textsuperscript{1}, Cassandra~Donatelli\textsuperscript{3}, Kayla~Hall\textsuperscript{4},\\ Adam~Summers\textsuperscript{4},
  Daniele~Panozzo\textsuperscript{1}, Cl\'audio~T.~Silva\textsuperscript{1}\\
 \affaddr{\textsuperscript{1}New York University; \textsuperscript{2}Link\"{o}ping University; \textsuperscript{3}Tufts University}; \textsuperscript{4}University of Washington}\\
 \email{\textsuperscript{1}\{francis.williams,harishd,panozzo,csilva\}@nyu.edu}; \email{\textsuperscript{2}alexander.bock@liu.se}; \email{\textsuperscript{3}cassandra.donatelli@tufts.edu}; \email{\textsuperscript{4}\{kchall8,fishguy\}@uw.edu}
}

\maketitle

\begin{abstract}
The \emph{ScanAllFish} project is a large-scale effort to scan all the world's 33,100 known species of fishes. It has already generated thousands of volumetric CT scans of fish species which are available on open access platforms such as the Open Science Framework. To achieve a scanning rate required for a project of this magnitude, many specimens are grouped together into a single tube and scanned all at once. The resulting data contain many fish which are often bent and twisted to fit into the scanner. Our system, \emph{Unwind}, is a novel interactive visualization and processing tool which extracts, unbends, and untwists volumetric images of fish with minimal user interaction. Our approach enables scientists to interactively unwarp these volumes to remove the undesired torque and bending using a piecewise-linear skeleton extracted by averaging isosurfaces of a harmonic function connecting the head and tail of each fish. The result is a volumetric dataset of a individual, straight fish in a canonical pose defined by the marine biologist expert user. We have developed Unwind in collaboration with a team of marine biologists: Our system has been deployed in their labs, and is presently being used for dataset construction, biomechanical analysis, and the generation of figures for scientific publication. 
\end{abstract}




\section{Introduction} \label{sc:introduction}
New tools often lead to scientific discoveries, and this is particularly true for new 3D imaging technology, which has helped advance many scientific areas. The availability of 3D imaging scanners has resulted in tens of thousands of large datasets to be analyzed. Our work is centered on the \emph{ScanAllFish} and \emph{oVert} projects, which are large-scale efforts to (CT) scan all the world's known species of fishes and other vertebrates~\cite{scan-all-fish-science, Watkins2018}. 
The Micro-CT scan allows scientists to determine the skeletal structure of a variety of fishes, opening the doors to a better understanding of relationships among skeletal elements and the degree of skeletal mineralization, as well as enabling population-wide studies that were previously impossible. Both volumetric and surface renderings are useful for making quantitative measures of skeletal parameters that are used to build evolutionary trees and demonstrate the directional variation of morphology over evolutionary time~\cite{Hall2018, Kolmann2018, Stocker2019}. 

The extracted skeletal geometry can be used to make physical models of function and support the understanding of swimming motions by combining finite element modeling and computational fluid dynamics. Finally, scans allow researchers, curators, and scientific communicators to make 3D printed replica of these fishes for expositions and museum archives.
The huge number of fish, their variety, and the scanning techniques involved cause unique challenges. As previously described in Bock~\cite{Bock:2018}, fishes are packed together for scanning purposes, with multiple fishes being placed inside a single scanning chamber. Every fish is twisted in a different way, they have different sizes and shapes, and those that are too long are \emph{curled up} to fit into the scanner. 
This method of packing multiple fishes together allows for rapid scanning of multiple species, but causes difficulty in analyzing the raw volumes. In fact, there are two fundamental problems that hamper effective use of the data: (1) the difficulty of separating each fish into its own volume and (2) dealing with fishes that are bent and twisted in different poses, making side by side comparisons impossible.

While the first problem can be addressed using existing segmentation techniques~\cite{Bock:2018}, the second problem is the challenge that we address in this paper: we propose an interactive pipeline to reverse this undesired deformation, restoring the original shape of the fish into a canonical straight pose, and thus facilitating the analysis and visualization of these valuable datasets. 

\hl{
Ideally, fish straightening would be performed completely automatically. Unfortunately, this requires the detection and measurement of the distortion that each exemplar underwent during the packing in the CT scanner. This information is impossible to acquire with a CT scan, since it requires the measurement of the volumetric stresses in the exemplar itself. 
Custom interactive tools therefore play an important role in helping users effectively guide this straightening process.
While existing off-the-shelf software support deforming 3D volumes, these approaches require users to work directly in 3D making its usage time consuming even for experienced users, let alone our target users who are not familiar with 3D modelling software.
Hence, it is essential to optimize user interactions so as to not overburden the users in their workflow. In fact, minimizing human effort for various tasks has recently gained traction in the design of user interfaces. For example,
Hong~et~al.~\cite{Hong2014} showed that designing an interface with minimal user-selectable information was most effective in the context of understanding accessibility in cartographic visualization; Ono~et~al.~\cite{Ono2019} designed an interface to track baseball plays that reduces the annotation burden on the user. Similarly, Choi~et~al~\cite{Choi2019} proposed an approach that automatically emphasizes words within a document and prividing recommendations in order to reduce the burden on users labeling documents.
}

Following along the above strategy, we design \oursystem, a user-driven, interactive volumetric straightening system that provides a single interface to start working directly with the original CT data, and enables the marine biologist to quickly and efficiently process the twisted volumes into clean data in a canonical straight pose. The user interaction is divided in two phases: (1) a selection phase, where the user picks a pair of 3D points to identify the extrema of the spine of a fish, from which the system automatically extracts a skeleton and an initial approximation of the straightened fish; and (2) a refinement (or tuning) phase, where the user navigates the 2D cross-section of the fish and fine-tunes the deformation by specifying additional rotations required to eliminate the torque and bending in the fish. 
The system is based on a simple but novel deformation method which is specifically designed for undoing the bent and twist introduced during the packing of multiple fishes, and that can be efficiently implemented on a GPU to ensure an interactive volumetric rendering of the undeformed dataset. \oursystem is already in use in the labs participating in the ScanAllFish project.

\hidecomment{
The contributions of this paper are:
\begin{itemize}
    \item A deformation algorithm and corresponding interactive user-interface to straighten CT datasets. The method provides an interactive, volumetric rendering of the undeformed fish, and allows marine biologist to process a dataset in 6 minutes on average.
    \item An open-source, reference system implementation.
    \item A demonstration of the effectiveness of \oursystem from expert feedback and a large collection of interactive sessions, whose videos are attached in the additional material.
\end{itemize}
}

The contributions of this paper can be summarized as follows:
\begin{itemize}
    \item The design of an interactive tool allowing users to process a CT image, extracting individual fishes, and straightening them. This tool allows a marine biologist to process a dataset in 6 minutes on average. %
    \item A simple deformation algorithm that enables an intuitive and easy user interaction by allowing users to interact with 2D slices instead of the complete 3D volume.
    \item A preliminary user evaluation, comparing the time and quality obtained by 10 users processing a representative dataset.
    \item A demonstration of the effectiveness of \oursystem through expert feedback on processing 18 fishes.
    \item A reference system implementation.
\end{itemize}

\section{Related Work} \label{sc:related}

Our approach combines techniques from geometry processing, to estimate the initial deformation, with rendering and deformations techniques developed within the visualization community. Here, we give an overview of the most closely related works, and we refer to~\cite{PMP:2010,Johnson:2004:VH:993936} for a complete overview.

\paragraph{Skeletonization via Discrete Maps.}
We review here the skeletonization works applicable in our setting, and we refer an interested reader to~\cite{Tagliasacchi:2016} for a detailed overview. 

Our skeleton construction is based on a harmonic volumetric parametrization \cite{Wang:2004:harmonic,Li:2007:Harmonic} constructed from a pair of user-provided landmarks. The isosurfaces of the scalar function are averaged to find points in the center of the fish: this construction is inspired by the hexahedral method proposed in \cite{gao2016structured} and the tubolar parametrization proposed in \cite{Livesu:2017}. 

It is important to observe that the parametrization induced by these functions is not bijective \cite{Schneider:2015}, and it is thus not a proper foliation \cite{Campen:2016}: however, this is not a problem in our case since we use it only to compute an approximate skeleton. We opted for this skeleton extraction procedure since it allows users to intuitively and interactively control the skeleton, which is mandatory to make our system able to process challenging datasets. For a complete overview of skeletonization techniques, we refer an interested reader to \cite{Tagliasacchi:2016}.

\paragraph{Volumetric Deformation.}
This problem has been heavily studied in both in the context of (1) \emph{volumetric parametrization}, where a energy minimizing, quasi-static solution is found via numerical optimization, (2) in \emph{physical simulation}, where the focus is on modeling dynamics effects, and (3) in \emph{space warping techniques}, where a explicit reparametrization of the space is used to warp an object. Since we are not interested in dynamics (we only need to deform the volume once), we only review the parametrization and free form deformation literature, and we refer an interested reader to \cite{Witkins:1997,Meier:2005,Nealen:2006,Sifakis:2012,skinningcourse:2014} for an overview of dynamic physical deformation techniques.

\paragraph{Volumetric Parametrization.} A convex approximation of the space of bounded distortion (and thus inversion-free) maps has been proposed in \cite{Lipman:2012,Kovalsky:2015}, allowing to efficiently generate these maps both in 2 and 3 dimensions. These methods do not require a starting point, but they might fail to find a valid solution in challenging cases. A different approach, guaranteed to work but requiring a valid map has been proposed in \cite{Hormann:2000,Degener:2003}: the idea is to evolve the initial map to minimize a desired cost function, while never leaving the valid space of locally injective maps. Many variants of this construction have been proposed, either enriching existing deformation energies with a barrier function \cite{Schuller:2013}, or directly minimizing energies that diverge when elements degenerate \cite{Smith:2015}. Specific numerical methods to minimize these energies have been proposed, including coordinate descent~\cite{Hormann:2000,Fu:2015}, quasi-newton approaches~\cite{Smith:2015,Kovalsky:2016,Rabinovich:2017,Shtengel:2017,Claici:2017}, and Newton \cite{Schuller:2013} methods. A last category of methods \cite{Fu:2016,Poranne:2017} produces an initial guess separating all triangles and rotating them into the UV space, and then stitches them together using Newton descent. However, all these methods are computationally intensive, and not suitable for interactively deforming high-resolution CT scans.

\paragraph{Space Warping.} Closed-form volumetric deformations have been defined using lattices  \cite{Sederberg:1986,Barr:1984,Coquillart:1990} or other parametrizations  \cite{Chen:2003}. While not directly minimizing for geometric distortion, they have the major advantage of being directly usable in a volumetric rendering pipeline, enabling to render in real-time the deformed volume. Our approach is also directly usable in a real-time volumetric rendering pipeline and uses a keyframed skeletal parametrization to define the deformation. 

Volume wires \cite{Volume_Wires} uses a skeleton to define a free form deformation, parametrizing it with values attached to the skeleton itself. Volume wires relies on a computationally intensive evaluation which prevents its use in a real-time rendering systems. Our method shares the idea of using a skeleton to parametrize the deformation, while providing detailed deformation control using keyframes, an algorithm to automatically estimates an initial deformation, and being specifically tailored to be used in an interactive volumetric rendering system.

\paragraph{Interactive Applications.} 
Many variants of the previous volumetric deformation techniques have been used in interactive applications in the visualization community: since a complete overview is beyond the scope of this work, we limit our review to the most closely related works, and we refer an interested reader to the surveys by Sun~\textit{et al}~\cite{sun2013survey} for visual analytics, and Liu~\textit{et al.}~\cite{liu2014survey} for information visualization techniques. 

Closely related to this work, Correa~\textit{et~al.}~\cite{correa2007volume} introduced an interactive visual approach to deform images as well as volumetric data based on a set of user-defined control points. Here, the deformation is controlled based on the movement of these control points. Even though their formulation introduced distortions in other regions of the data, since their focus was on illustrative applications and volume exploration and visualization, such distortions were acceptable since they were occluded in the visualization. On the other hand, our goal is to generate data that will be further analyzed by the marine biologists. It is therefore necessary that such distortions are avoided.
Such distortions are common in other approaches as well that perform volume deformation with the focus on exploration and/or animations~\cite{correa2007illustrative,Kwon:2018}.

There have also been visual approaches that target generating illustrations with the focus on medical data~\cite{MRH08,Li:2007}, in particular, generating views when the covering surface is ``cut open". Since these approaches distort the data that is deformed, they are not suitable for our work.
Nakao~\textit{et~al.}~\cite{Nakao:2010,Nakao:2014} proposed an interactive volume deformation, also catered towards medical applications, which deforms the volume based on a proxy geometry that approximated the volume. However, the proxy geometry itself is computed in a preprocessing phase, which makes the combined pipeline not interactive. 

Directly related to CT scans of fishes, in our previous work we proposed TopoAngler~\cite{Bock:2018} that combines a topology based approach with a visual framework to help users segment fishes from the CT data. TopoAngler is used as a preprocessing step in this work, to extract a segmented fish (Section \ref{sec:system}).

\section{Unwind: Design Overview}

Processing one CT dataset containing a packed set of fishes requires multiple steps, which are currently only possible by using different software packages, and that require conversions and manual processing to be combined: (1) to process the scanned images into a 3D volume format (and optionally subsample the volume), (2) to segment and export individual fishes from the 3D volume, and (3) to deform the fishes into a canonical pose. %
While straightening the individual fishes is desirable, given that there exists no off-the-shelf tool to accomplish this, it was not possible for the users to do this.
Our goal in the design of \oursystem is to provide a single, efficient, and intuitive tool to process the CT data once they obtain it from the scanning software, enabling non-expert users to process the massive amount of data which is acquired daily in marine biology labs. To accomplish this, we divide the entire process into four stages:

\noindent \textbf{1. Load CT data:} The user can directly load the output from the CT scanner, which is then subsampled to fit in the video memory of the workstation to ensure an interactive preview of the deformation. %
    
\noindent \textbf{2. Segment and extract a single fish:} For the initial segmentation, we decided to integrate the functionality from the open sourced TopoAngler~\cite{Bock:2018} due to three reasons: (1) it is widely used by marine biologists, being the only tool specifically designed for this tasks; (2) the user interaction is intuitive, requiring only a single parameter accompanied with a few clicks from the user to select the required subvolumes and, (3) it is interactive, providing a live preview even on complex volume scans of multiple fish.
This segmentation approach first generates a hierarchical segmentation based on the join tree of the input data~\cite{CSA03}, and then allows the user to interactively choose the simplification to be performed and select the segmented sub-volumes corresponding to the fish.
We refer the reader to \cite{Bock:2018} for a detailed description of this algorithm.

\noindent \textbf{3. Estimate the straightened volume:}
In this stage, our goal is to semi-automatically estimate the straightened volume using user input. Since the spine (or the mathematical skeleton) of the fish traces out a curve in space, it was a natural decision to define our deformation as a warping of a curve in space. This requires to compute the spine geometry followed by defining the deformation based on the warping that straightens it. 
While automatically extracting the spine for a given fish would be ideal, existing skeletonization methods are not easily adaptable in our setting given the varied shapes and sizes different fishes take. Therefore, we decided to employ a minimalistic user assisted approach, wherein we require the user to select the two extremal end points of the fish, which are then used to find the fish skeleton. Note that the alternative of allowing the user to manually specify the 3D curve corresponding to the spine is an arduous task requiring the interaction with a 3D volume.

\noindent \textbf{4. Refine the straightening.}
Since our deformation occurs along a curve in space, it was very natural to allow the user to interact with 2D cross sections along that curve.
Once the approximate spine is computed, we define a set of coordinate frames along this curve, which is then used by the user to refine the deformation. This process mainly involves the user aligning the coordinate frames to generate an accurate spine from the estimated curve.
Using such an interface was inspired by popular video editing software in which users can edit a set of keyframes. The simplicity of editing in 2D combined with real time 3D visualization of the deformation makes for an easy-to-use tool allowing extremely precise control over the deformation. Not only is this deformation transformation simple mathematically (thus allowing interactivity), but users could also easily understand this procedure simply by using the software without requiring a mathematical explanation.

The next two sections focuses on the third and fourth steps of the above workflow, and describes in detail the user interface of \oursystem. 
We would like to note that the described system was designed over multiple iterations spanning over a year based on constant feedback from our collaborators (who were also the initial users). We discuss this process after the description of the user interface. 

\section{Cylindrical Deformation}
\label{sec:space}
\begin{figure}[t]
    \centering
    \includegraphics[width=\columnwidth]{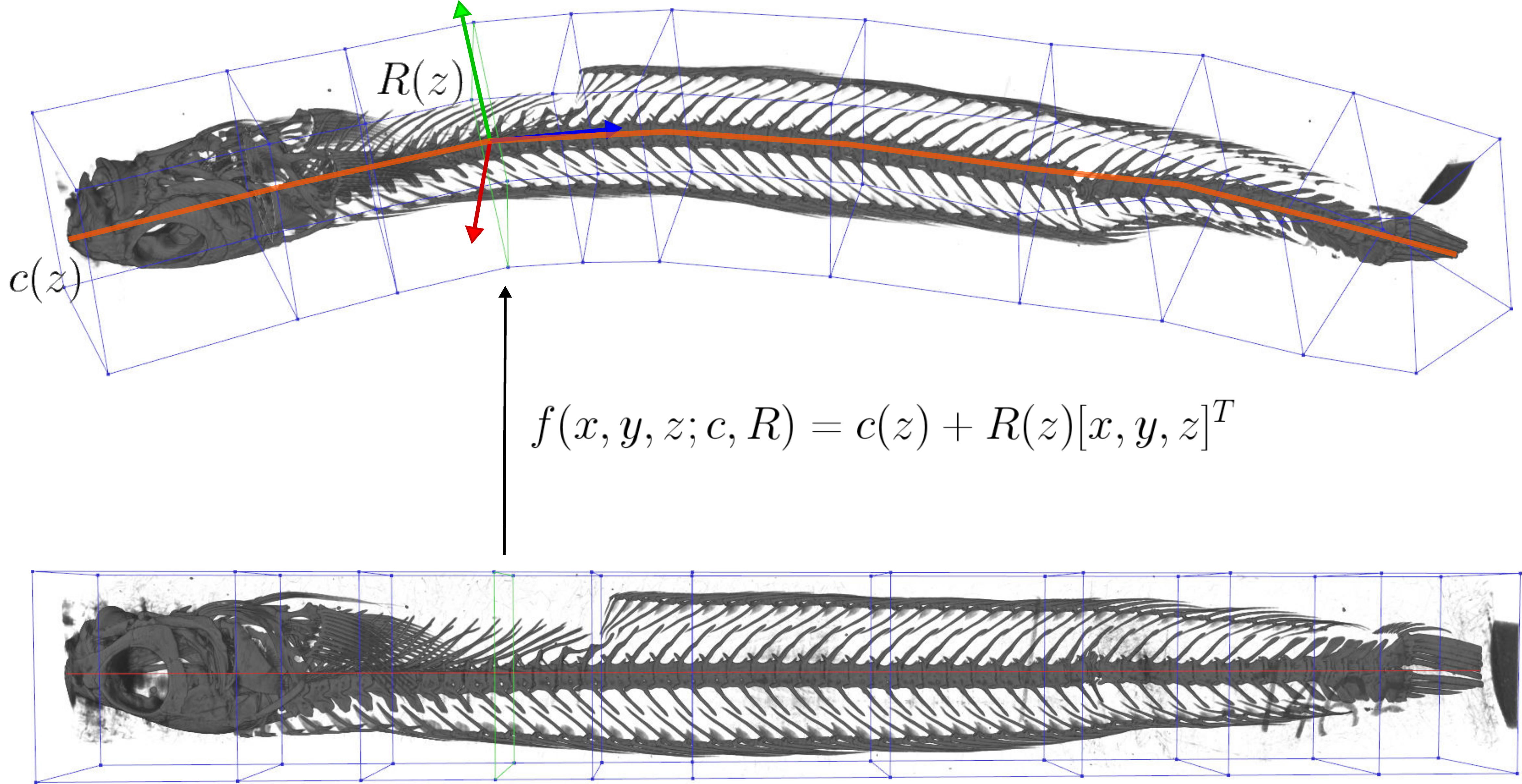}
    \caption{An illustration of our volumetric deformation: The density at a point $(x, y, z)$ in the straight volume is determined by the cross sectional plane $R(z)$ centered at $c(z)$, where $c(z)$ is a piecewise linear curve sweeping the spine of the fish and $R(z)$ is a frame centered at $c(z)$ defining the orientation of the fish at a given position. $R(z)$ and $c(z)$ are specified at a set of \emph{keyframes} (the blue squares) and interpolated linearly in between.}
    \label{fig:fish_volume_deformation}
\end{figure}
\newcommand{\vtwist}{V_\text{twist}}
\newcommand{\vstraight}{V_\text{straight}}
\newcommand{\finv}{f^{-1}}
\newcommand{\R}{\mathbb{R}}

Without loss of generality, we assume that fishes are
straightened one at a time. If more than one fish is present in the scanned volume, we isolate individual fishes using TopoAngler~\cite{Bock:2018}.
The key idea in our approach is to identify a deformation function $\f$, which transforms an axis aligned bounding box (which will contain the straight fish), into a deformed version of the fish. This function, being the inverse of the deformation that the fish underwent, will be then used to recover the straight fish.

More concretely, we are interested in a mapping, $\f$, which deforms a straight cylindrical region $\vstraight$ into a deformed one $\vtwist$ such that the central axis of the $\vstraight$ is mapped to a curve $c(t)$, which is aligned with the body of the twisted fish. Furthermore, for every $z$ coordinate in $\vstraight$, we define a rotation matrix $R(z)$ which maps points off the central axis to points in $\vtwist$.
The set of rotation matrices $R$ captures both the ``bends" that the fish underwent as well as the ``twists" (or torsion) in the fish.
Figure~\ref{fig:fish_volume_deformation} illustrates one such deformation function $f$.

Thus, to parameterize $f$, we require a space curve $c(t) = \begin{bmatrix}c_x(t), c_y(t), c_z(t)\end{bmatrix}^T$ and a continuous field of rotation matrices $R(t) = \begin{bmatrix}u(t)^T, v(t)^T, n(t)^T\end{bmatrix}^T \in SO^{3}$, allowing us to write:
\begin{equation}
    \label{eq:cylindrical}
    \f(x, y, z; c, R) = c(z) + x \cdot u(z) + y \cdot v(z)
\end{equation}
Intuitively, the direction along $n$ in $R$ points along the skeleton (central axis) of the fish, while the $u$ and $v$ directions define the right and up directions of the fish respectively. Thus, $n$ allows us to undo any bending in the fish while $u$ and $v$ allow us to remove torsion. 

In our setting, we use an arclength parameterized piecewise linear curve for $c(t)$. Note that the arclength parametrization ensures that the mapping will be close to an isometry independently on the speed of the parametrization of $c$. This is important since it allows the user to freely change the parametrization speed by adding additional keyframes, without introducing unwanted distortion (Section \ref{sec:system}). Specifically, $c(t)$ can be parameterized by vertices $e_1, \ldots e_m \in \R^3$. Letting $d(e_i) = \sum_{j < i} ||e_{j+1} - e_j||_2$, we can write $c(t)$ explicitly as:
\begin{equation}
    c(t) = \lambda(t) e_{k(t)} + (1 - \lambda(t)) e_{k(t)+1}
\end{equation}
where 
\begin{align}
    k(t) = \argmin_{i, (d(e_i) < t)} (t - d(e_i))\\
    \lambda(t) = \frac{t}{e_{k(t)} - d(e_{k(t)+1})}
\end{align}

To define $R(t)$, we define orthonormal coordinate frames $R_1 = (u_1, v_1, n_1)^T, \ldots, R_m = (u_m, v_m, n_m)^T \in SO^{3}$ at each vertex $e_i$ of $c$. For $R(t)$ to be continuously defined at all points along $c$, we identify the unit normals, $n_i$, with points on a sphere and spherically interpolate the $n$ directions between adjacent $R_i, R_{i+1}$:
\begin{equation}
    R(t) = \text{SLERP}(n_{k(t)}, n_{k(t)+1}, \lambda(t)) R_{k(t)}
\end{equation}

As we show next, the parameters of the deformation function $f$ are initially estimated by our system, and optionally interactively refined by the user in a second stage to compute the final straightened fish.

\section{Interactive Fish Straightening} 
\label{sec:system}

\oursystem uses the cylindrical deformation (Equation~\ref{eq:cylindrical}) to assist users in straightening deformed fishes. Once the individual fish is isolated from the input CT data, the remaining stages of the process can be further divided into the following 4 steps:
\begin{enumerate}
    \item Compute the harmonic function used for estimating the deformation function $f$. 
    \item Estimate the parameters of the deformation function $f$.
    \item (Optional) Refine the parameters of the deformation function $f$.
    \item Export the straightened fish.
\end{enumerate}
We now describe in detail each of these steps and the associated visual interface of our system. Each step comprises of its own set of visualization widgets, and the user can move forward and back between the different steps. The entire workflow is illustrated in the accompanying video. Note that the user input is used to guide this process during the different steps. 

\hidecomment{
\subsection{Identify Sub-Volume}
As mentioned earlier, the input to our system is a 3D volume (structured volumetric grid) containing the scan of a single fish. Due to the long scanning time, the ScanAllFish project participant always scan tens of fishes in the same scanning volume: we use topology-based segmentation tool TopoAngler~\cite{Bock:2018} to extract single exemplars.
This approach first generates a hierarchical segmentation based on the join tree of the input data, and then allows the user to interactively choose the simplification to be performed and select the segmented sub-volumes corresponding to the fish.
We refer the reader to \cite{Bock:2018} for a detailed description of this process.
}

\subsection{Compute Harmonic Function}
\label{sec:harmonic}

We first approximate the fish by a smooth curve in order to estimate the parameters of the deformation function $f$. This curve is computed as the set of centroids of level sets of an harmonic scalar field defined on the volume, following a technique similar to \cite{gao2016structured}. 

\paragraph{Discretization.} Different techniques could be used to compute the harmonic function, and we opted for a finite element method due to its efficiency, simplicity, and robustness. While it is possible to use directly the voxel grid as a space discretization, this would be prohibitively expensive on the high resolution CT scan. Downsampling the image is a possibility, but it would lose the high-frequency details and risk to lead to disconnected components in the thin regions of the fish. We therefore opt for an adaptive tetrahedral mesh, generated using an implementation of Isosurface Stuffing~\cite{labelle2007isosurface}, which strikes a good balance between boundary approximation and computational efficiency.
While unlikely, it is possible that the generated tetrahedral mesh is made of multiple disconnected components, either due to lack of resolution, or due to the TopoAngler segmentation. To ensure there is only a single connected component, we inflate the voxel grid until all the connected components are merged using \cite{Chen:2018}.

\paragraph{User-Provided Extrema.}
\begin{figure}[t]
    \centering
    \includegraphics[width=\columnwidth]{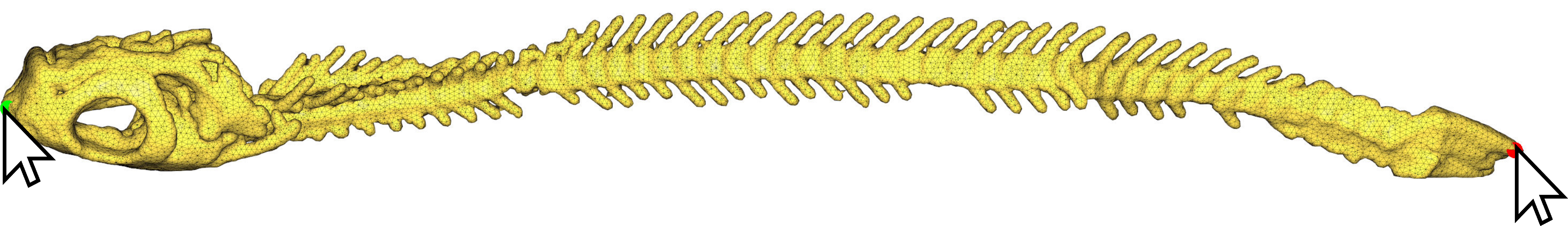}
    \caption{The user selects two extrema on the endpoints of the fish by clicking on the extracted tetrahedral mesh.}
    \label{fig:2clicks}
\end{figure}
The extrema of the harmonic function, used as boundary conditions, are provide by the user with an end-point selection widget allowing the user to select points on the segmented fish (see Figure~\ref{fig:2clicks}). These points corresponds to the head and tail of the fish, making it straightforward for the user do this selection.

The two endpoints are then used to compute a discrete harmonic function with Dirichlet boundary conditions setting the head vertex $\vtxhead$ and the tail vertex $\vtxtail$ as a source and sink:
\begin{equation}\label{eq:laplace}
    (Lu)_i = \begin{cases} 
    0  & \text{if } u_i \neq \vtxhead, u \neq \vtxtail \\
    1  & \text{if } u_i = \vtxhead \\
    -1 & \text{if } u_i = \vtxtail
    \end{cases}    
\end{equation}
Here, $L$ is the discrete Laplace-Beltrami Operator \cite{PMP:2010}, and $u$ is a scalar field defined at each vertex of the tetrahedral mesh. While solutions to Equation~\ref{eq:laplace} produce fields whose level sets trace out a reasonable skeleton approximation, the spacing between level sets is not uniform. To remedy this issue, we resample the curve using a fixed spacing between vertices in ambient space. To ensure the resampling process does not discard details, we use a sampling width which is half the size of the smallest segment in the curve traced out by the level sets of the harmonic solution.

\begin{figure}
    \centering
    \includegraphics[width=\columnwidth]{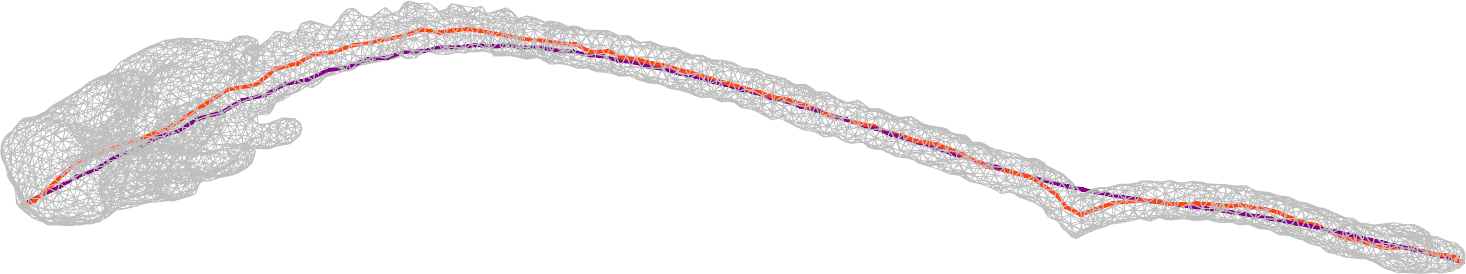}
    \caption{Estimated skeletons using the centroids of level sets: The \emph{red curve} shows the piece-wise linear curve where the vertices are 100 centroids of the level sets uniformly sampled on the scalar function $u$ (Equation~\ref{eq:laplace} defined on the tetrahedral mesh. The \emph{purple curve} is the result of applying 50 smoothing steps the red curve.}
    \label{fig:noisy_skeleton}
\end{figure}

\subsection{Estimating Parameters $c(t)$ and $R(t)$}
\label{sec:estimating_params}
To generate an initial estimate for the piecewise linear curve, $c(t)$, we sample $k$ level sets of $v$ at isovalues uniformly spread on $[-1, 1]$. We then compute the centroids, $c_1, \ldots, c_k \in \R^3$, of these level sets. While the piecewise linear curve whose vertices are the $c_i$'s traces a curve approximating the bend of the fish, the curve itself might be noisy due to the complex boundary geometry. We thus apply $s$ iterations of Laplacian smoothing (replacing every vertex with the average of its 2 neighbours) to smooth the curve. Figure~\ref{fig:noisy_skeleton} compares two curves before and after smoothing. Specifically, if $c_{i-1}, c_{i},$ and $c_{i+1}$ are consecutive vertices of the curve, one iteration of smoothing can be written as:

\begin{equation}
    \text{SMOOTH}(c_i) = c_i^\prime = \frac{c_{i-1} + c_{i+1}}{2}
\end{equation}

Once we have vertices $c_i^\prime$, we then compute orthogonal coordinate frames $R^\prime_1, \ldots R^\prime_k$, where $R_i^\prime = (u_i^\prime, v_i^\prime, n_i^\prime)^T \in \R^{3 \times 3}$. First we compute $n_i^\prime$ at each of the $c_i^\prime$ using central differences on the interior and one sided differences at the boundary:

\begin{equation}
    n_i^\prime = \begin{cases}
    \frac{c^\prime_{i+1} - c^\prime_{i+1}}{||c^\prime_{i+1} - c^\prime_{i+1}||_2} & \text{if } 1 < i < k \\
    \frac{c^\prime_{i+1} - c^\prime_{i+1}}{||c^\prime_{i+2} - c^\prime_{i}||_2} & \text{if } i = 1 \\
    \frac{c^\prime_{i+1} - c^\prime_{i+1}}{||c^\prime_{i} - c^\prime_{i-2}||_2} & \text{if } i = k
    \end{cases}
\end{equation}

\begin{figure}
    \centering
    \includegraphics[width=\columnwidth]{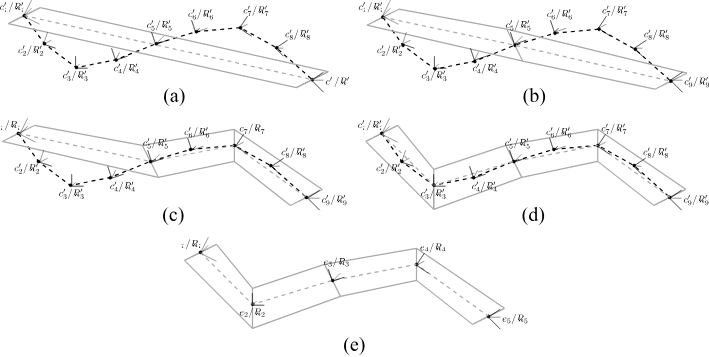}
    \caption{Computing a minimal set of parameters by subdivision: The black dotted line illustrates the initial estimated parameters $c_i^\prime$ and $R_i^\prime$, using the method described in Section~\ref{sec:estimating_params}. We compute the gray prism by connecting two squares on the $u_1^\prime-v_1\prime$ and $u_9^\prime-v_9^\prime$ planes. We progressively subdivide the prism at vertices $c_i^\prime$ until the subdivided prisms fully contains all the $c_i^\prime$. This refinement procedure yields a new, reduced set of parameters $e_1, \ldots e_5$ and $R_1, \ldots, R_5$.}
    \label{fig:prism_subdiv}
\end{figure}

We then compute $u_i^\prime$ and $v_i^\prime$ by projecting the $x$ and $y$ axes into the plane defined by $n_i^\prime$. If such a projection degenerates, we repeat this procedure with the $x$ and $z$ axes as well as the $y$ and $z$ axes (one of them has to succeed since the skeleton is not degenerate by construction): 

\begin{align}
    u_i^\prime = \hat{x} - \hat{x}^T n_i * n_i,\\
    v_i^\prime = \hat{y} - \hat{y}^T n_i * n_i.
\end{align}

The result is a piecewise linear curve with vertices $c_1^\prime, \ldots c_k^\prime$ and orthonormal bases $R_1^\prime, \ldots R_k^\prime$ at each vertex. Note that the number of level set samples, $k$, and smoothing iterations, $s$ are user-tunable parameters. The default is $k=100$ and $s=50$ and our users did not change them in any of their experiments. 

\paragraph{Computing a Minimal Set of Parameters.}
Having a large numbers of parameters can become cumbersome to a user when refining the deformation (see Section~\ref{sec:refine} below). Thus, while we could use the parameters $c_1^\prime, \ldots c_k^\prime$ and $R_1^\prime, \ldots R_k^\prime$ for the deformation, we opt instead to compute a minimal set of parameters $e_1, \ldots, e_m$ and $R_1, \ldots R_m$, $R_i = (u_i, v_i, n_i)^T$, which agree with the $c_i^\prime$'s and $R_i^\prime$'s. 

To compute these new parameters, we select a radius $r$ and construct a prism whose bases are squares with side lengths $2r$. Each base is centered at $c_1^\prime$ and $c_k^\prime$ and is oriented to lie in the planes $n_1^\prime$ and $n_k^\prime$ with the sides aligned with $u_1^\prime$, $v_1^\prime$ and $u_k^\prime$, $v_k^\prime$. Figure~\ref{fig:prism_subdiv}(a) shows a 2D illustration of the initial configuration for an example curve.

Then, while the prism does not fully contain the vertices, $c_1^\prime, \ldots c_k^\prime$, we subdivide it by first choosing a vertex $c_\text{mid}^\prime$ and frame $R_\text{mid}^\prime$, and then splitting a prism into two with a base centered at $c_\text{mid}^\prime$ and aligned with $R_\text{mid}^\prime$. Figure~\ref{fig:prism_subdiv}(b)--(e) illustrates this subdivision procedure.

The resulting $e_i$ and $R_i$ are the vertices and coordinate frames of the subdivision location used to construct the prism. The radius hyperparameter, $r$ is user selectable: A larger radius will yield a coarser approximation, and a smaller radius will yield a finer one. By default we set $r$ to 10 voxel-widths, and our users did not adjust it in any of their experiments. 

\begin{figure}
    \centering
    \includegraphics[width=\columnwidth]{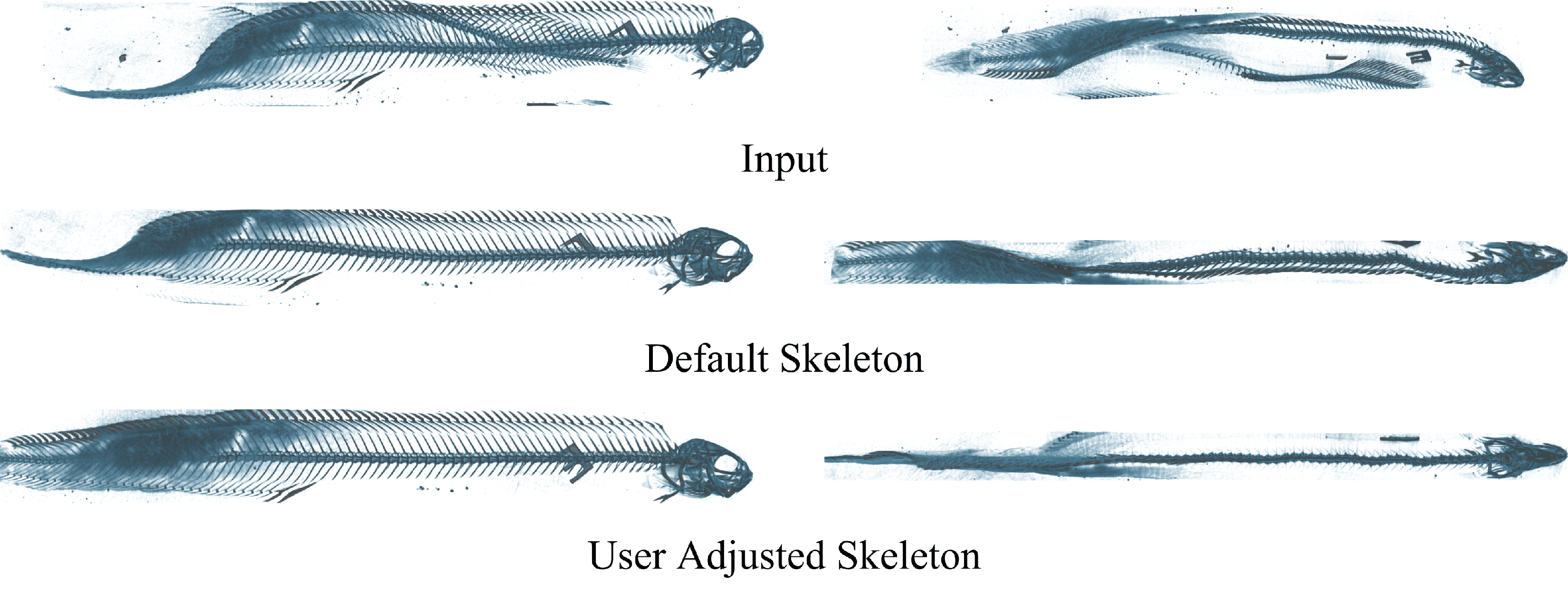}
    \caption{The deformation automatically estimated from the input (top) captures the majority of the distortion (middle) and can be further refined adding additional keyframes (bottom).}
    \label{fig:automatic}
\end{figure}

Figure~\ref{fig:automatic} shows an example of the initial estimated deformation on a fish scan. 

\paragraph{Handling Disconnected Components.}
\begin{figure}[t]
    \centering
    \includegraphics[width=0.7\columnwidth]{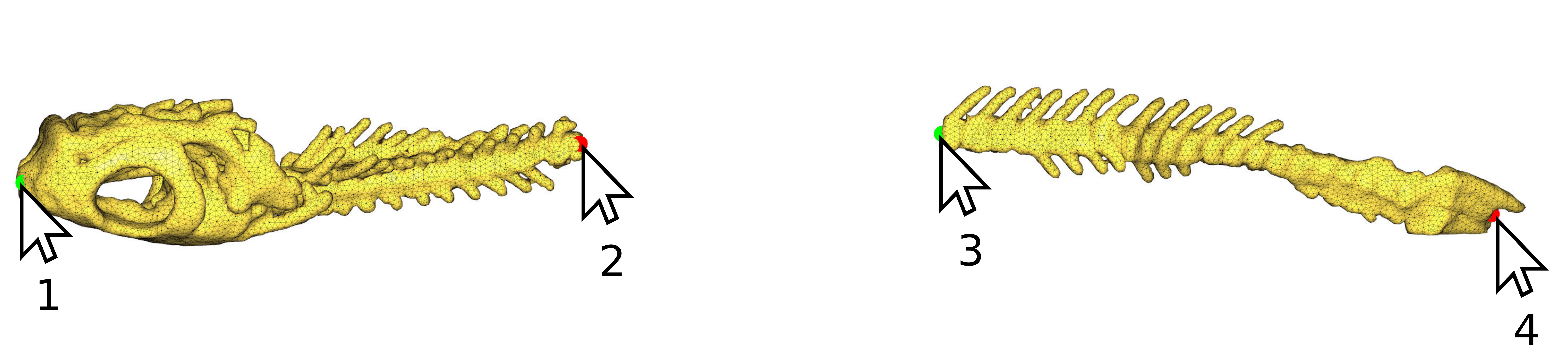}
    \caption{In cases where the segmentation outputs disconnected components, we recover a curve approximating the skeleton by selecting endpoints on each connected component. In the figure, the user selects points 1, 2, 3, and 4 and the system estimates a curve between 1 and 2 and another between 3 and 4 which are joined by a line connecting 2 and 3.  }
    \label{fig:4clicks}
\end{figure}
There are cases where the components corresponding to a fish are significantly far apart that the dilation operation performed when identifying the sub-volume corresponding to the fish is not sufficient to merge the two components (see Section~\ref{sec:feedback} for details).
To handle such a scenario, we allow the user to select the end points of the different components in order, compute the harmonic function and estimate the deformation function parameters for each component, and use the ordered end points to merge the different curves into a single one (Figure \ref{fig:4clicks}). 

\subsection{Deformation Refinement}
\label{sec:refine}

\begin{figure}
    \centering
    \includegraphics[width=\columnwidth]{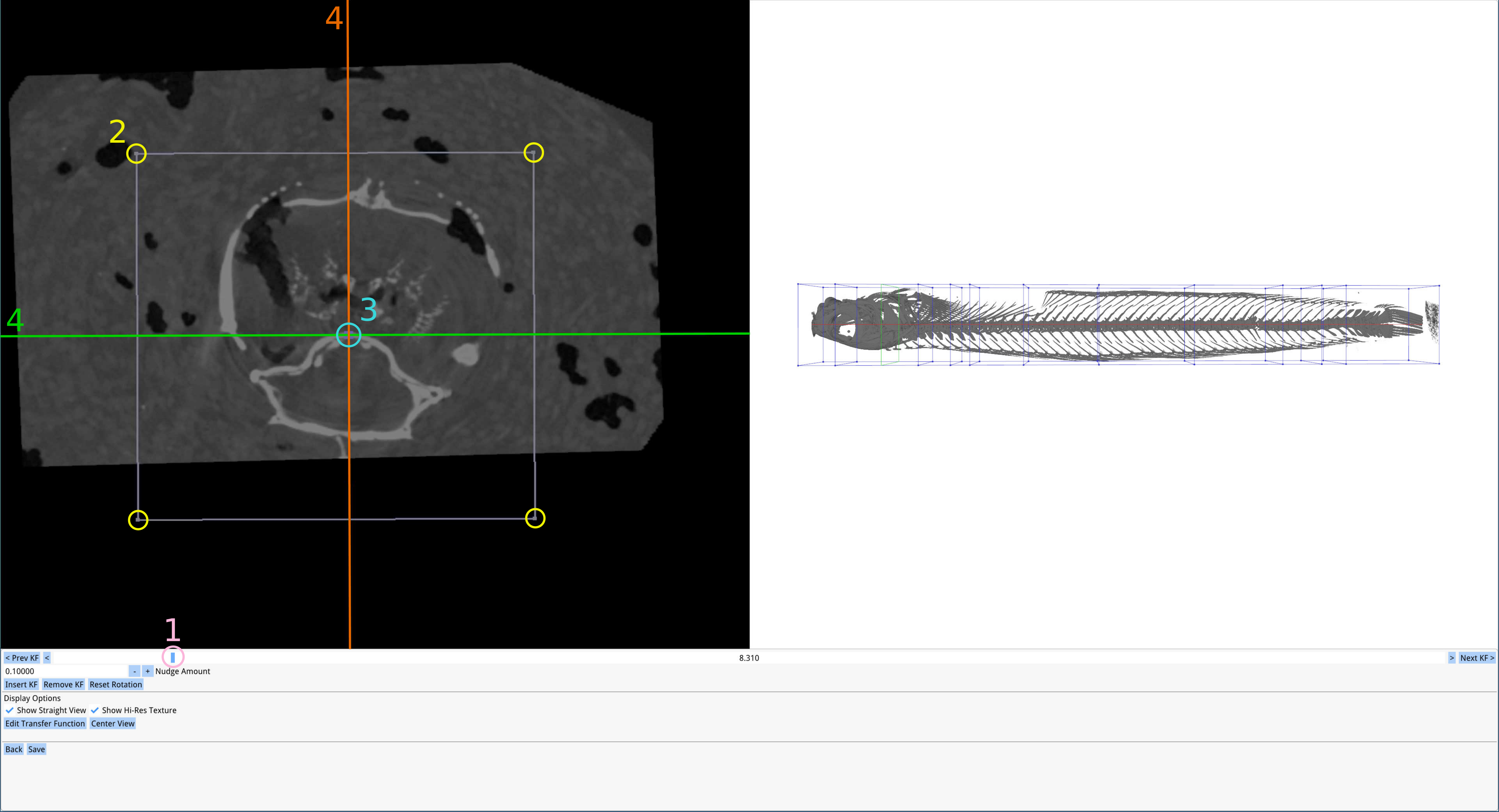}
    \caption{Our user interface for performing interactive deformations.
    }
    \label{fig:UIcontrols}
\end{figure}

We allow the user to interactively edit and refine the deformation parameters. The user interface for this step comprises 3 widgets (Figure~\ref{fig:UIcontrols}): a \textit{2D view} for editing in a cross section (Figure~\ref{fig:UIcontrols}, left), a \textit{3D view} showing the effect of the edits in real time (Figure~\ref{fig:UIcontrols}, right), and a \textit{control widget} providing buttons and sliders to help the user perform the deformation (Figure~\ref{fig:UIcontrols}, bottom).

\paragraph{Control Widget.}
This widget provides the user with a slider (Figure~\ref{fig:UIcontrols}, 1) that controls the parameter $t$ along the parametric curve $c(t)$. Users can use this to move along this curve to identify locations to edit the deformation. Buttons at either end of the slider allow the user to skip to parameters corresponding to the parameter vertices, $e_i$. 
The widget also provides options for the user to add new vertices, $e_i$ and frames, $R_i$, as well as delete existing vertices and key frames.
Alternatively, any interaction in 2D view will automatically add a new vertex and key frame along the curve.
Additionally, this widget enables the user to view and change the transfer function of the rendered volume and includes controls to recenter the camera and toggle between the straight and deformed views.

\paragraph{2D View.}
This view shows a cross section of the volume in the plane orthogonal to $n(t)$ corresponding to the currently selected parameter $t$ in the control widget. Visual cues are overlaid over this cross section to allow user modify the different deformation parameters. These cues include:
\begin{enumerate}
    \item A box corresponding to the region of space in the input which will be deformed to generate the output (Figure~\ref{fig:UIcontrols}, 2). This represents the the prism base corresponding to the current parameter vertex $e_i$. The user can adjust the size of the bounding prism, $V_\text{cage}$, by dragging the corners of this box. 
    \item A point corresponding to the position of $c(t)$ in the 2D plane orthogonal to $n(t)$ (Figure~\ref{fig:UIcontrols}, 3). Dragging this point allows the user to change change the position of vertices on the parametric curve.
    \item The directions $u(t), v(t)$ in the 2D plane. The user can also  rotate the $u(t)$ and $v(t)$ vectors around $n(t)$ by holding shift and dragging (Figure~\ref{fig:UIcontrols}, 4). This feature allows the user to align $u(t)$ and $v(t)$ with the principal directions of the fish, thus allowing for the removal of any torsion which may be present in the twisted input.
\end{enumerate} 

\paragraph{3D View.}
Depending on the option selected in the control widget, the 3D view visualizes in real time, either the straightened fish or the deformed bounding cage and curve $c(t)$. The former option allows the user to receive real-time feedback on how their edits affect the output, while the latter view allows the user to visualize how well their curve approximates the skeleton of the fish.
The straightened volume is obtained by sampling the deformation function $f$ after each update to its parameters. This is accomplished by mapping a regular lattice (which is used for the volume rendering) to the input volume which is then sampled using trilinear interpolation. To provide real time feedback at interactive rates, we perform this sampling using the fragment shader as part of the rendering pipeline.   

\subsection{Real-Time Rendering and Export}
\label{sec:export}

Our system, similarly to existing volume deformation pipelines \cite{Chen:2003}, enables real-time rendering of the warped volume, allowing the user to instantaneously see the result of their actions. To render deformations in real time, our system computes a straight volume by evaluating $f(x, y, z)$ (Equation~\ref{eq:cylindrical}) along cross sections in $z$. We implement this evaluation in an OpenGL shader which renders cross sections along $z$ into a volume texture. The texture size is determined by the prisms described in Section~\ref{sec:estimating_params}. 

Once the user is satisfied with modeled deformation, the straightened volume can be exported onto disk. We use the same procedure for exporting as we do for real time rendering. To preserve the correct dimensions during export, the depth ($z$-direction) of the volume is set to the arclength of the linear curve $c(z)$. The width and height ($x$ and $y$ directions) are set to maintain the same aspect ratio as the prisms. The user can optionally edit the size of the exported volume.

In addition to exporting the volume, our application allows the user to save a session to disk and reload it later for further editing or inspection. This feature is important to ensure \emph{provenance}, enabling to store a direct mapping between the straightened fish and the original RAW volumetric dataset.

\hl{\section{Design Process}}
\hl{
Once the problem (straightening scans of fish) was identified, we examined example scans and analyzed the existing workflow used by the marine biologists for manipulating CT scans. 
While there are manual deformation systems built into several commercial programs (e.g., Amira, Aviso) which have been used in the past (e.g., straightening the deformed jaw of a megamouth shark~\cite{Dean2012}), these methods require users to manually place and adjust reference points in 3D to perform the necessary deformation. To quote one of our collaborator, when asked how effective these tools were for straightening CT scans, his response was: \emph{``I would say it simply cannot be done. A poor job took hours and hours over several days when I worked on the megamouth shark"}. 
}

\hl{
Moreover, the gold standard for 3D data acquisition demands that a museum specimen be imaged. That means the shape of the fish is set by the fixative used when the specimen was collected. So, fish are always, whether scanned singly or in groups, scanned with bent spines.
Manual, physical straightening of the fish before scanning is difficult, time consuming, and can lead to damage to the specimen, and it is thus not a viable option.
}

\hl{
So, our next step was to consider the adaptability of existing free form deformation based approaches to perform the deformation. This requires users to manually deform a lattice bounding the fish directly in 3D. This was an onerous task even for an expert savvy with 3D modelling tools and it requires a high learning curve. We therefore decided to build a custom tool for this purpose with aim of making the straightening process easy for our target users. In particular, our goal was to design an interface that requires minimal  and simple user interaction using metaphors that the users were already familiar with.
}

\hl{
The first version of our tool used a skeletal-based deformation performed by setting the length of the spine and the number of skeleton vertices. Here the user selects two endpoints on a segmented mesh, inputs a number of skeleton vertices, and the software computed a deformation using a volumetric extension of ARAP \cite{ARAP_modeling:2007} to map the automatically computed skeleton to a straightened mesh. 
While this approach had only a few user inputs, it often failed if the skeleton extraction was imperfect. It also required the segmented mesh to include minute details of the fish which could not be easily obtained through Topoangler.
}

\hl{
So, in the next iteration, we tried using a cage based deformation that used the above estimated skeleton to derive the initial bounding cage. The user then had to manipulate cage vertices to get the deformation. 
Again, depending on the quality of the segmented mesh obtained from Topoangler, the initial cage would often be far from the desired cage and hence required several interactions from the user to rectify it. Moreover, manipulating individual skeleton vertices was not only cumbersome, taking a lot of user time, but it also involved a high learning curve especially for users not familiar with 3D modeling tools.
}

\hl{
To overcome the problems caused by the coarse segmentation, we decided to use our curve-based approach, which requires computing only the main spine of the fish. 
After observing the current tools used by the biologists for segmentation, we noticed they were split into a 2D editing widget for selecting a boundary along a cross section and a 3D widget for selecting an axis aligned cross section along \textit{x}, \textit{y} or \textit{z} axis. Given this familiarity, we decided to use a 2D cross section view as well to help users adjust the alignment of the estimated spine with the actual spine and used a 3D view to show the results of the adjustments in real time. 
This design also reduced the user interactions, and removed unnecessary input such as cage vertices. Furthermore, this design was also robust in the sense that even in rare cases when the skeleton estimation is far from the actual spine of the fish, the user can easily recover by fixing the spine vertices.
}

\hl{
The above version of the tool was deployed at our collaborators' labs during which time we were actively collecting feedback and fine tuning the system. In particular, as more people started using the tool for straightening different fish scans, certain important shortcomings were noticed that had to be fixed. In particular,
(1)~there were cases where the segmented fish consisted of disconnected segments, which had to be fixed (as described earlier);
(2)~the users requested the ability to rotate keyframes along a second axis (in the initial version one could only rotate about the normal tangent to the skeleton curve). This was necessary to correct minor warping that was sometimes caused in the output volume; and
(3)~Since the estimated skeleton tracks the spine, in some cases when the straightened volume was exported, fleshy parts of the bits near the endpoints of the spine were missing. To overcome this, an option was added to pad the exported volumes with additional keyframes at the endpoints.
}

\section{Results} \label{sc:results}

The results reported in this Section were generated over a period of 8 weeks in the \emph{Friday Harbour Laboratory}~(FHL) in the Biology Department at \emph{University of Washington}. Users used Unwind to straighten fishes on a workstation with a Intel Xeon CPU E5-1607 v4 @ 3.10GHz, 32 GB RAM and a NVIDIA GTX-1080-Ti GPU. In the supplemental material, we include raw screencasts of the editing sessions for all the results presented in the paper. Our reference implementation and one sample dataset can be downloaded from our GitHub website \url{https://github.com/fwilliams/unwind}: binaries are also provided for Windows and Linux.

\paragraph{Synthetic Evaluation.}
\begin{figure}
    \centering

    \includegraphics[width=\columnwidth]{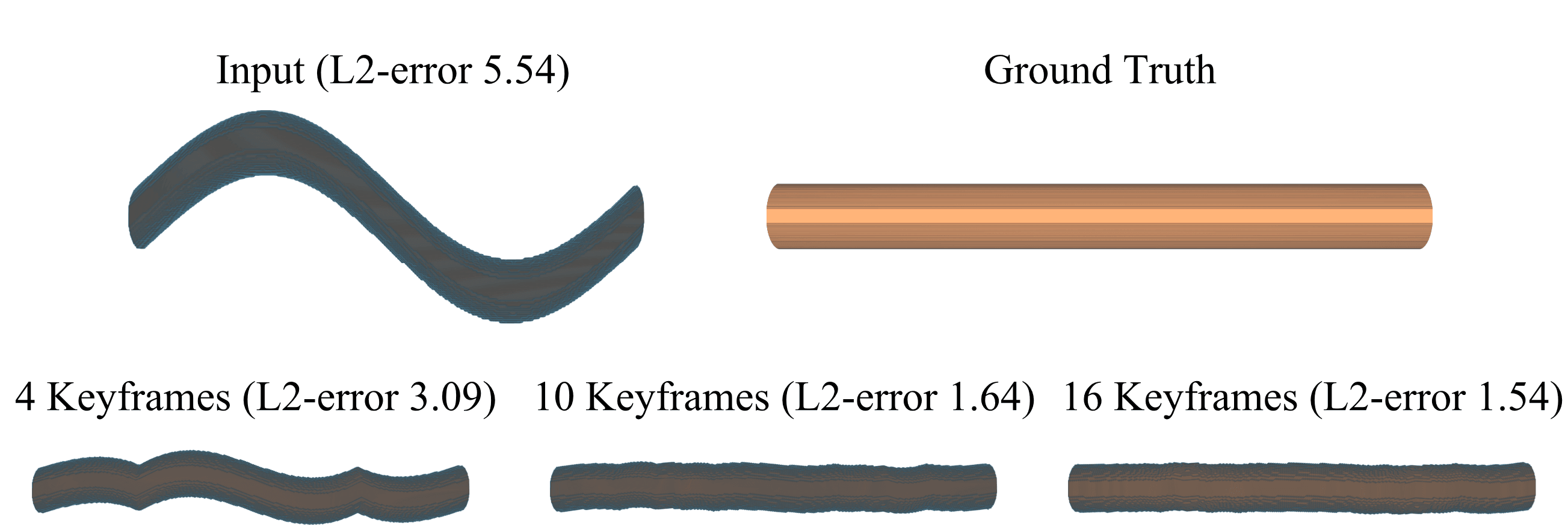}
    \caption{Straightening a synthetic example of a cylinder bent into the shape of a sine wave. As the user adds new keyframes, the result more closely approximates the ground truth cylinder. 
    }
    \label{fig:sine_wave_cylinder}
\end{figure}

\begin{figure}
    \centering
    \begin{scriptsize}
    Input Scan\\
    \includegraphics[width=0.5\columnwidth]{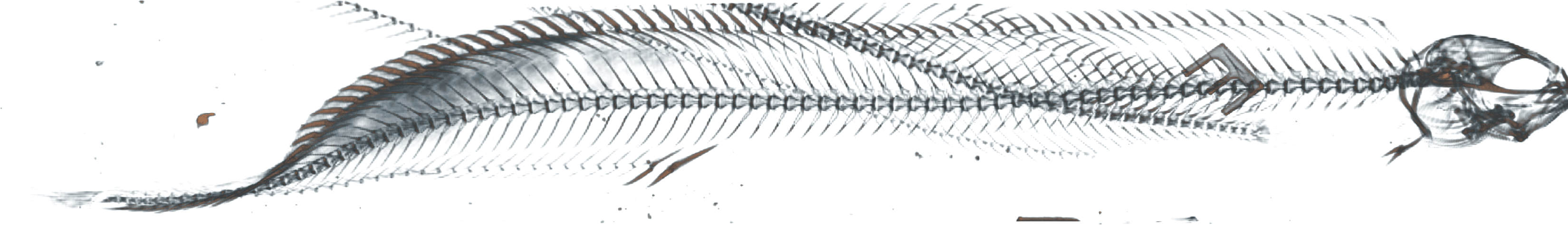}\\
    
    Results\\
    \includegraphics[width=0.4\columnwidth]{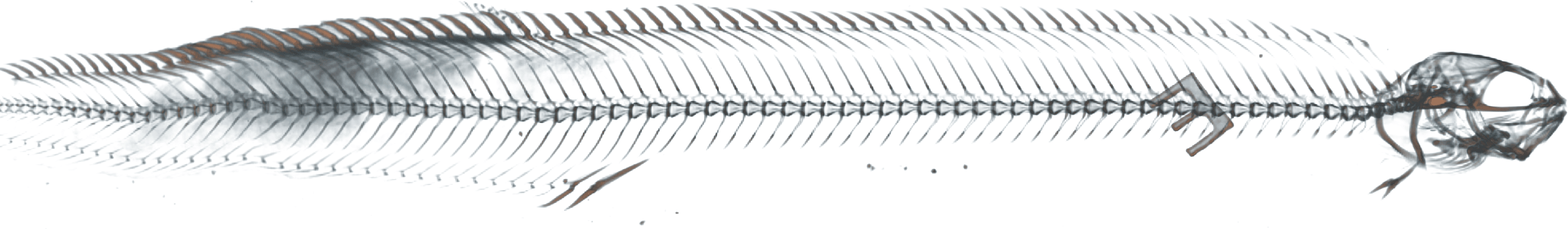}
    \includegraphics[width=0.4\columnwidth]{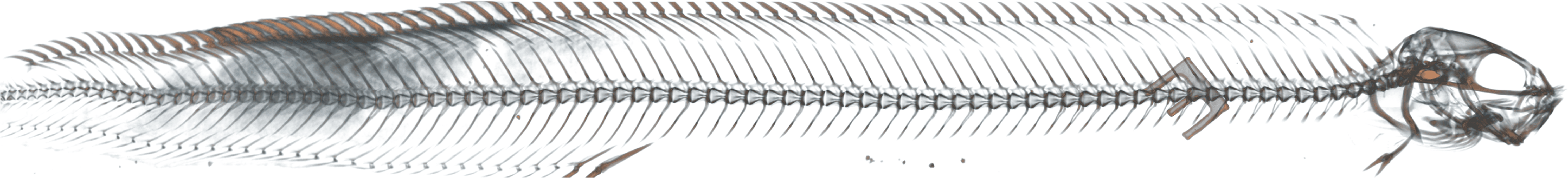}
    \includegraphics[width=0.4\columnwidth]{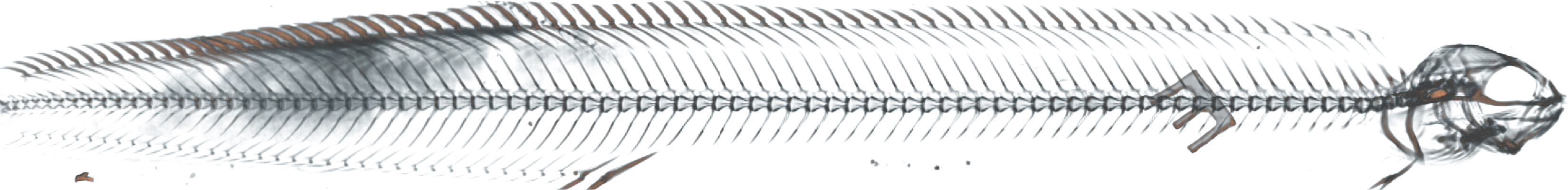}
    \includegraphics[width=0.4\columnwidth]{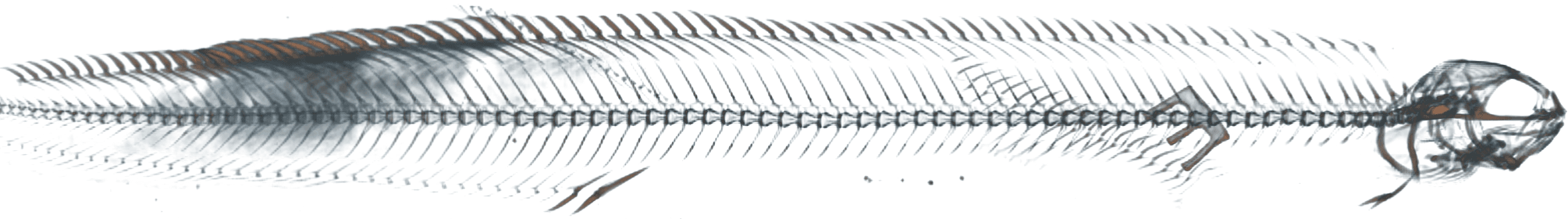}
    \includegraphics[width=0.4\columnwidth]{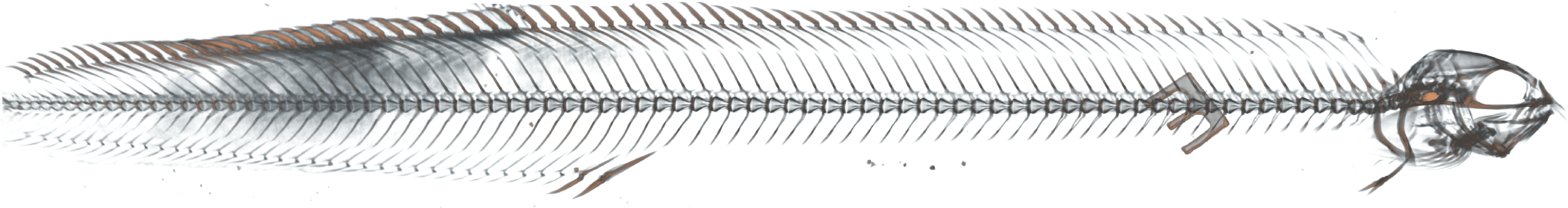}
    \includegraphics[width=0.4\columnwidth]{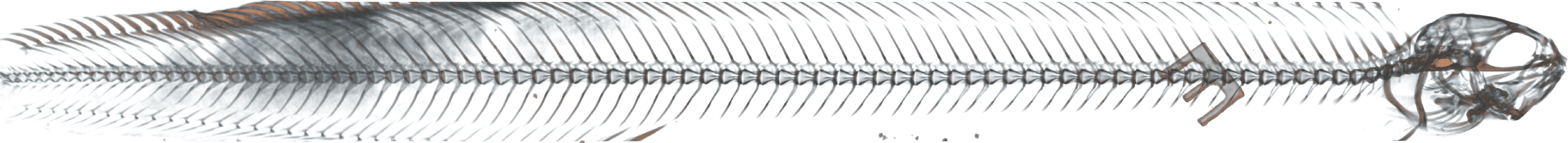}
    \includegraphics[width=0.4\columnwidth]{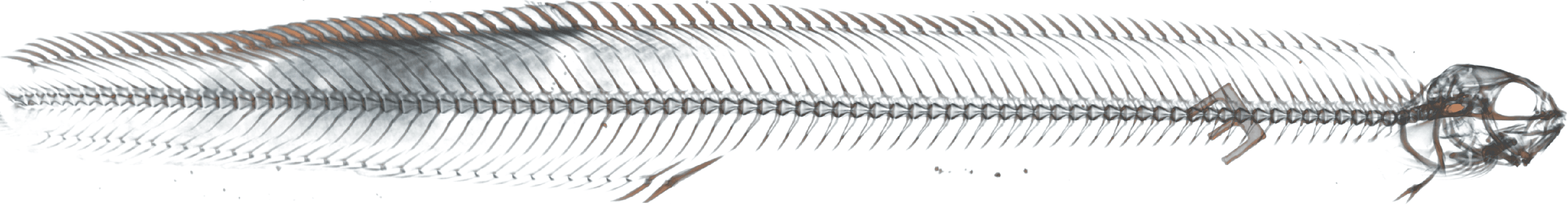}
    \includegraphics[width=0.4\columnwidth]{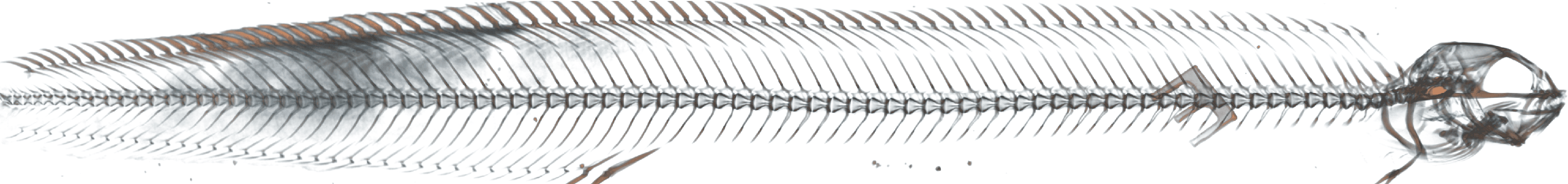}
    \includegraphics[width=0.4\columnwidth]{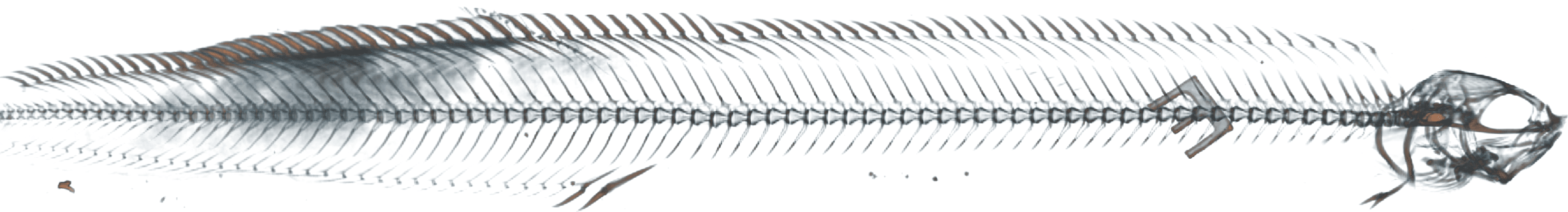}
    \includegraphics[width=0.4\columnwidth]{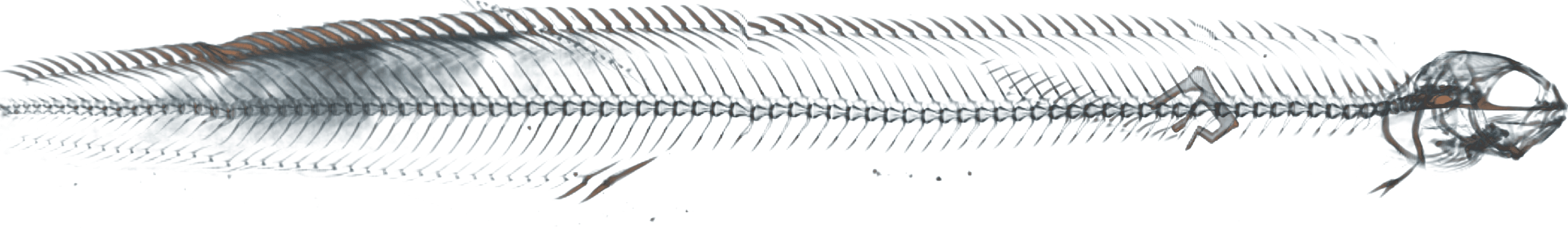}
    \end{scriptsize}
    
    \caption{The same fish straightened by 10 different users.}
    \label{fig:10fish}
\end{figure}

\begin{table}
    \centering
    \begin{tabular}{lrr}
         \textbf{Phase} & \textbf{Average Time}& \textbf{\%} \\
         \hline
         \\
         File Selection & 0m19s & 5.3\% \\
         Data Loading and Contour Tree & 1m27s & 24.4\% \\
         TopoAngler Segmentation & 0m29s & 8.1\% \\
         Mesh Extraction & 0m03s & 0.1\% \\
         Endpoint Selection & 0m07s & 1.9\% \\
         Straightening & 3m30s & 58.8\% \\
        \\
         User Interaction & 4m27s & 74.8\% \\
         Processing & 1m31s & 25.2\% \\
         \\
         Total & \textbf{5m57s} & 100\% \\
    \end{tabular}
    \vspace{0.5em}
    \caption{Average time taken by one expert user took to straighten 15 fishes. The top part of the table shows the time taken in each phase as well as loading times between phases. The bottom part shows how much of the time was spent by the user and now much time the system spent performing computations.}
    \label{tab:user_timings}
\end{table}

\begin{figure}
    \centering
    Input\\
    \includegraphics[width=0.9\columnwidth]{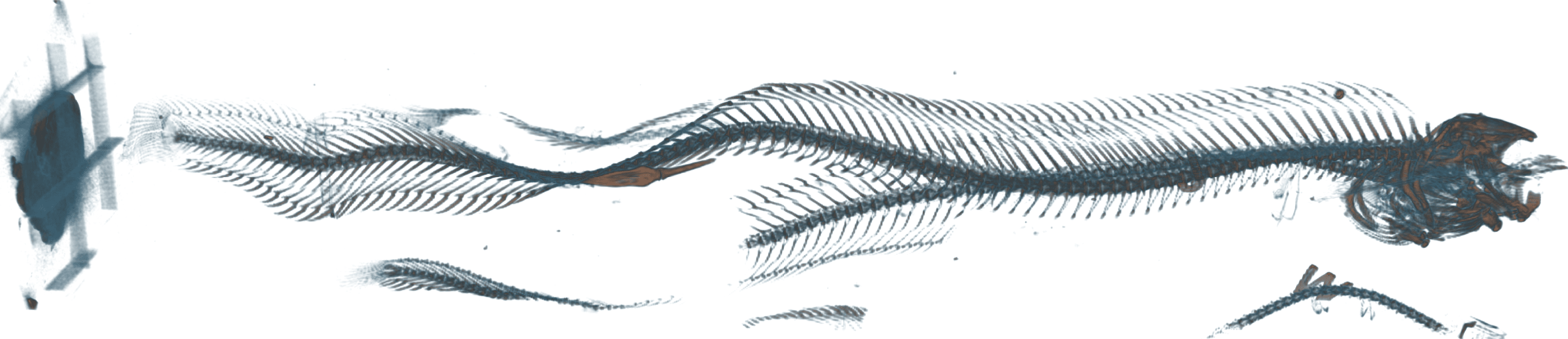}\\
    \includegraphics[width=0.9\columnwidth]{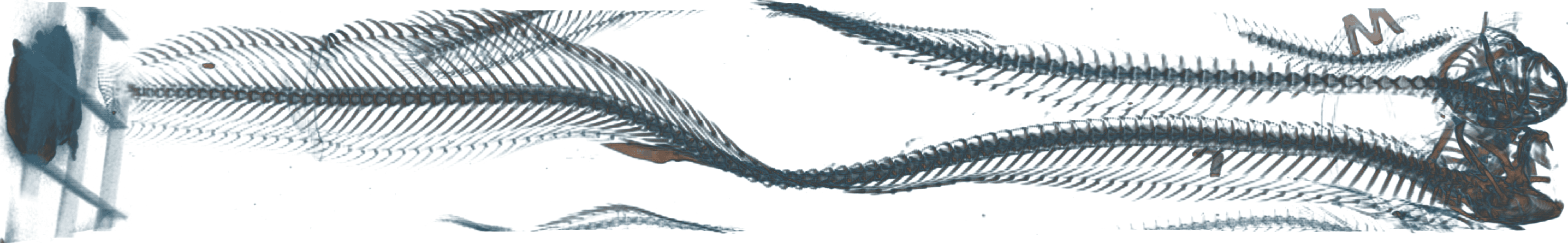}\\
    Output\\
    \includegraphics[width=0.9\columnwidth]{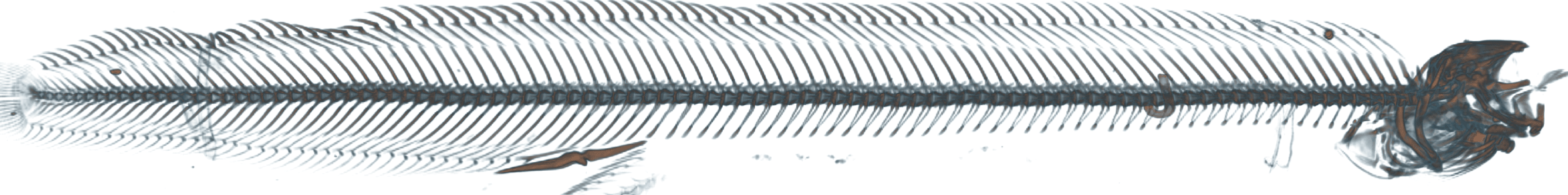}\\
    \includegraphics[width=0.9\columnwidth]{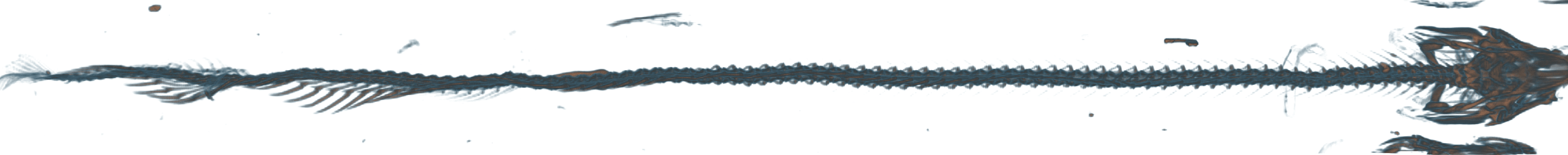}\\
    \caption{This scan was purposefully twisted to evaluate the performance of our system. In spite of the fact that the scan contains large twists as well as multiple fishes, our system produces a straightened volume for a single exemplar.}
    \label{fig:fish05}
\end{figure}

We performed a synthetic experiment to verify that our system works on analytic deformations. In Figure \ref{fig:sine_wave_cylinder}, we show how a cylinder whose medial axis is a sinusoid can be deformed back to a straight line. We use a different number of keyframes, to show the effect of the refinement on the final deformation. We also compute the $L^2$ distances (normalized by the length of the ground truth volume) between the reconstructed volume and the ground truth. 

\paragraph{Baseline Comparison.}
\begin{figure}
    \centering
    \includegraphics[width=\columnwidth]{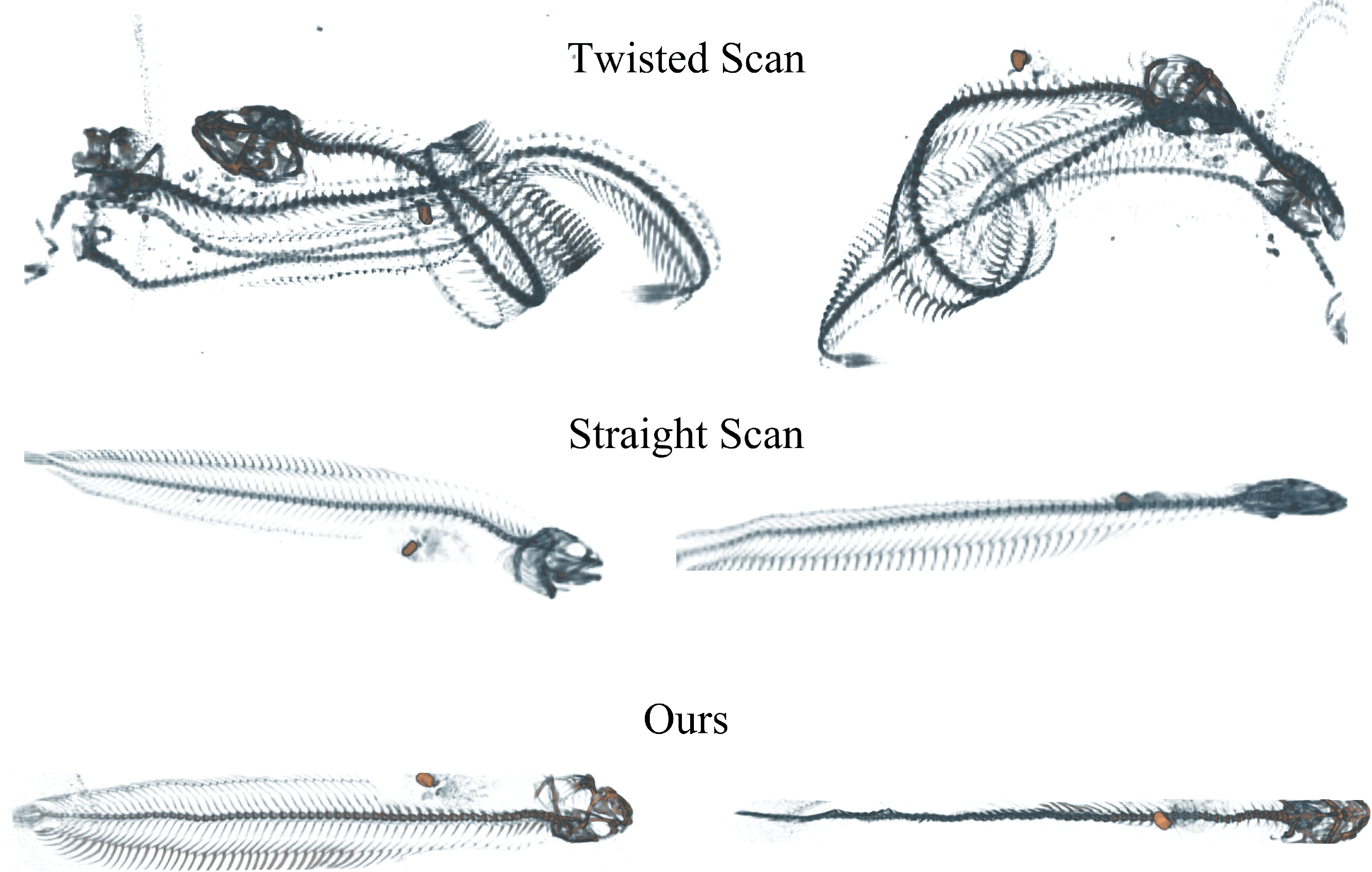}
    \caption{Here we scan a fish in both a twisted (top) and straightened (middle) pose. We compare the straightened scan with the scan produced by our method.}
    \label{fig:comparison}
\end{figure}
The motivation for our project is the avoidance of the practical hurdle of scanning each fish individually in a CT scan, enabling to scan tens of fishes in the same session. Since our algorithm deforms the raw data acquired by the scanner, we want to evaluate the possible artefacts and errors introduced in the results. To quantify these effects, we performed the following experiment: we scanned a fish individually in a straight pose, and scanned the same fish in a bent pose when acquired together with other fishes. We then compared the raw straight scan, with the untwisted fish created with our algorithm. The two results are similar (Figure \ref{fig:comparison}), with our result being more straight, since it is extremely challenging to physically straighten (and keep in a stable position) the fish for the entire duration of the scan. This result indicates that our software solution is an ideal replacement over individually scanning straight fishes, since it provides superior quality and massively reduces the scanning time.

\paragraph{User Experiment 1: Single CT dataset processed by multiple users.}
To evaluate our system, we selected a representative CT scan, and processed it in 3 different ways: (1) An expert user used a combination of existing tools, using TopoAngler to segment and Blender to deform using FFD, (2) the same expert user processed the dataset with our algorithm, and (3) a group of 10 novice users (marine biologists that never used our system before) processed it using our system. The 10 users were asked to approximately match the result produced by the expert user using FFD. 

We attach a video of (1) and (2) in the supplementary material, and we show the 10 straightened fishes by novice users in Figure \ref{fig:10fish}. The expert user has ample experience with 3D modeling software and practiced the task for 3 times both for FFD and with our software. He spent 7 minutes and 36 seconds for FFD, and only 1 minute and 33 seconds with our system. The novice users were trained in a 10 minutes overview of the software, and then spent between 6 and 19 minutes, with an average of 12.5 minutes, to complete their task. This indicates that our system provides a massive speedup over baselines for expert users, while also allowing novice users to be productive in this task with minimal training. Note that for these experiments we only included the time required to define the deformation function, since it is the only fair timing that we can use to compare different methods. Loading times, export times, and volumetric surface extraction were not included since they were done with different software stacks.

\paragraph{User Experiment 2: Untwisting a tank of fishes}
Our system is interactive: The large majority of the time is spent in the segmentation phase, and in the generation of the tetrahedral mesh and solution of the linear system to compute the skeleton. To quantify the timings of the different phases we asked one marine biologists to process a large CT volume containing 18 fishes. A breakdown of the timings is provided in Table \ref{tab:user_timings}. The overall (user + processing) average processing time per fish is around 6 minutes. These timings include the entire processing, and to put them in context with the acquisition pipeline it takes an experienced biologist 30 minutes to prepare, label, photograph, and pack the tube for scanning, and another 4-12 hours for the actual scan to complete. 
Our system thus adds a minor overhead to the overall acquisition pipeline.

\begin{figure}
    \centering
    \includegraphics[width=\columnwidth]{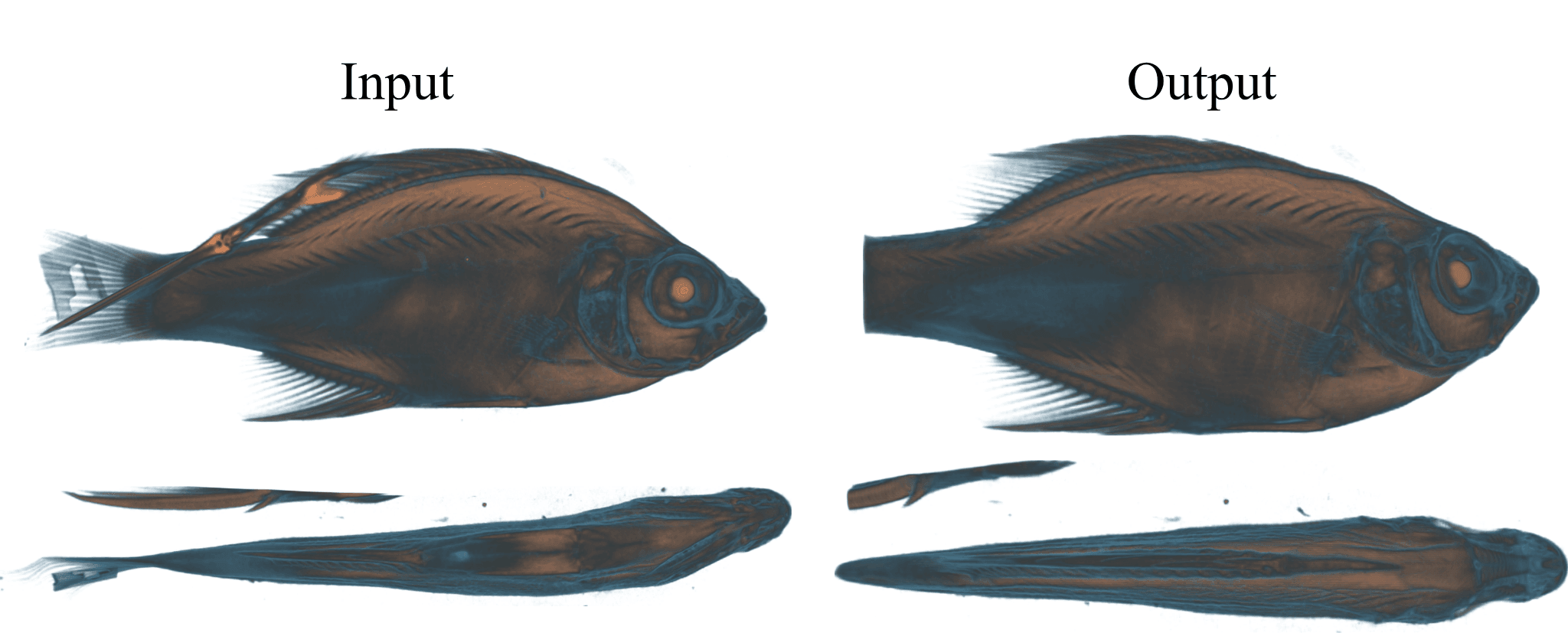}
    \caption{Our system can handle a wide variety of fish geometries.}
    \label{fig:fish06}
\end{figure}

\paragraph{Showcase of Results.}
Figures \ref{fig:fish05} and \ref{fig:fish06} demonstrate the capabilities of our system under a variety of challenging situations.

\begin{figure}
    \centering
    \includegraphics[width=\columnwidth]{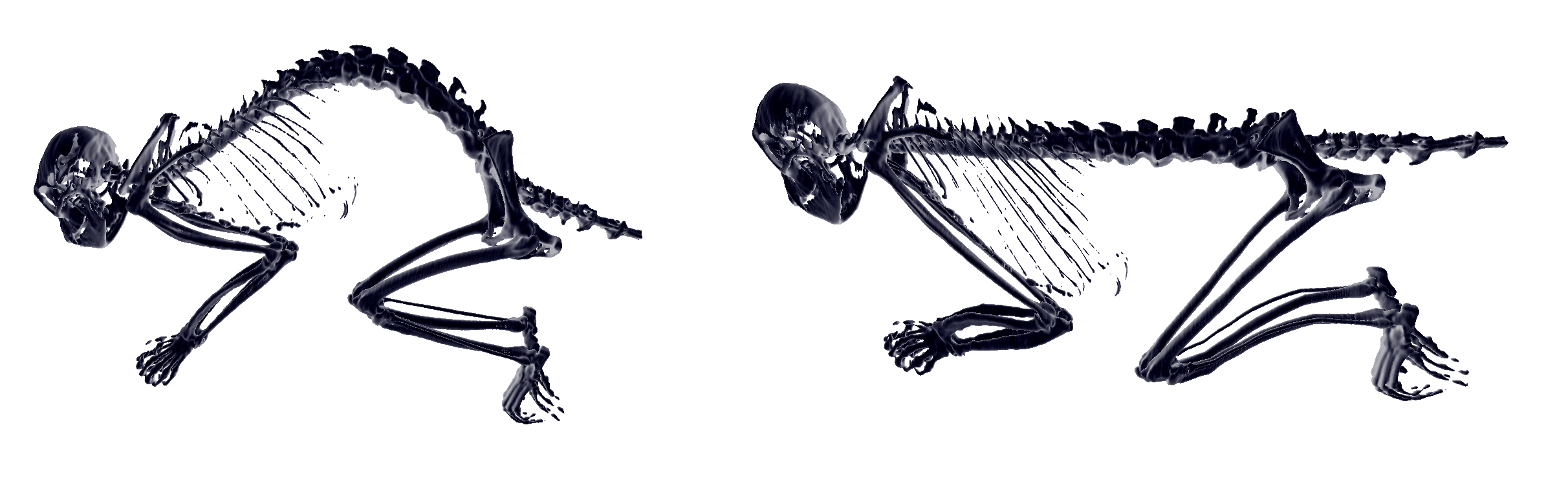}
    \caption{Example where the spine of a primate was straightened.}
    \label{fig:monkey}
\end{figure}

\subsection{Application of \oursystem on species beyond fishes.}
\hl{
Note that the initial target application of \oursystem was for straightening fishes. More than 40 people in the CT scanning community have so far been trained on using \oursystem at the FHL CT scanning facility. Currently, five of them are using it routinely and the data are being uploaded to MorphoSource.org as open access data sets. 
However, since we made our tool public, it has gained tremendous response from the community and has already been used to straighten and deform other species of animals for scientific analysis. Specifically, it has been used to straighten and measure spines of small primates, snakes, eels, and the "flaps" of stingrays.
Figure~\ref{fig:monkey} shows an example where the spine of a primate (Chlorocebus Aethiops) was straightened. 
}

\subsection{Expert Feedback}
\label{sec:feedback}

The overall feedback we received from our collaborators was positive with
them finding the software useful, intuitive and easy to use compared to existing off-the-shelves softwares. In fact, it took one of our collaborators only ``\textit{5 to 8 minutes}" to explain the working of the tool to a student, who was then able to use it independently. This has now become a staple application for them and has been integrated as part of their daily workflow. 

One of the primary advantages they found with straightening fish was when doing morphometrics analysis, especially when trying to look at the fish in a specific anatomical plane. As one of our collaborators mentions: ``\textit{Before, we had to have versions of the scan rotated at different angles to measure things in a somewhat straight line. Now we can just straighten a single scan which saves a lot of time, storage space, and confusion with different versions of files}". Our collaborator went on add that in addition to the analysis, our tool also significantly helps save time when preparing results for communication as the following quote testifies: ``\textit{it makes creating figures a lot easier. We don't have to spend hours looking for the perfect scan, we can just fix any one we have}".

However there were some difficulties that were discovered as the software was being used. 
Cases were discovered where there was a significant gap between parts of the fish, due to which the dilation performed was not sufficient to create a single connected component.
Such cases are possible if the skeleton of a fish is not necessarily connected (i.e., they might be connected via other tissues that have lower density than the bone), of the scanned fish is broken. We then had to then tweak both our interface, as well as the way the central axis of the fish is computed to handle such scenarios (see Section~\ref{sec:harmonic}).
There were also some interface requests, in particular, regarding the supported file formats. Currently, our software supports importing data from a stack of .bmp files, which is the output from the CT scanning software, and exports the straightened fish as a raw binary image. Given that users can also have processed data in other formats (tiff stack, DICOM stack, or nrrd files, etc.), having the having the ability to handle these formats will make it easier and faster to work with the tool. 

Based on one of the suggestions for next steps, we will be looking into also doing the  reverse---bend a fish to specific angles. This can be greatly useful for a variety of studies as well. For example, marine biologists also work with scans of fish with muscle fibers stained. In such cases, they would like to take a straight fish, and then see what the difference in muscle fiber length would look like in a bent fish. 

We attach the unmodified transcript of our interview in the additional material.

\section{Limitations and Concluding Remarks} \label{sc:conclusion}

We introduced a novel system for supporting the creation of a large encyclopedia of CT scans of fishes, providing a user-assisted procedure to undo the unwanted deformation introduced in the scanning process. We demonstrated the utility of our system, evaluating quantitatively the error introduced in baseline comparisons, and qualitatively over real-world scans. The system is available as open source, and it is in active use in our collaborator's labs.

\paragraph{Limitations.} Our system has a high GPU memory requirement, which requires a high-end graphics workstation to run. Reducing this requirement is important to foster the applicability of this system and enable it to be used by the community at large. Another limitation of our system is that for exemplars with high distortion, the initial deformation estimation can be inaccurate, requiring more user interaction than for other exemplars (Figure \ref{fig:automatic}). 
Much of the distortion is a result of the assumption that there is a straight line between the fish's snout and its tail. Fish skulls are highly complex and extremely diverse across species. One way to fix these large deformations would be to define the skull of the fish prior to the de-warping process. If we constrain slices between the front and back of the skull, it is likely to prevent most of the distortion and improve the efficiency of the workflow.

\paragraph{Future Work.} We believe that, after several months of usage, we will have access to sufficient data to replace the original deformation estimation with a data-driven model, trained on the data manually processed by the users. We are also porting our application to WebAssembly, to run directly in a web browser and making it easier to deploy in biological labs.

An additional, and surprising, application for this system is strategically bending fish that were scanned straight. Many scientists research fish muscle morphology and how local shape change during locomotion contributes to swimming kinematics. To do this, they stain specimens with iodine before scanning so that the muscle fibers are radio-opaque just like the skeleton. A modified version of Unwind could be developed to warp scanned fish to a shape they might attain during swimming, and look at the change in length of the muscle fibers.

\vspace{-0.5em}
\section{Acknowledgments}
\begin{footnotesize}
The authors would like to acknowledge Thomas O'Mahoney for the contribution of the primate scan in Figure~\ref{fig:monkey}.
This work was partially supported by the Moore-Sloan Data Science Environment at NYU; NASA; the NSF CAREER award 1652515; NSF awards OAC-1835712, CHF-1908767, NSF IIS-1901091, CNS-1229185, CCF-1533564, CNS-1544753, CNS-1730396, CNS-1828576; CNS-1626098, a gift from Adobe Research; a gift from nTopology; a gift from AMD, and the NVIDIA NVAIL at NYU. APS was supported by the Seaver Institute and NSF DBI-1759637. C.~T.~Silva is partially supported by the DARPA D3M program. Any opinions, findings, and conclusions or recommendations expressed in this material are those of the authors and do not necessarily reflect the views of DARPA. We would also like to acknowledge our trial users Kelly Diamond, Jonathan Huie, Karly Cohen, Todd Clardy, and Graham Short.
\end{footnotesize}

\bibliographystyle{SIGCHI-Reference-Format}
\balance
\bibliography{99-bib,topology}


\begin{thebibliography}{00}


\ifx \showCODEN    \undefined \def \showCODEN     #1{\unskip}     \fi
\ifx \showDOI      \undefined \def \showDOI       #1{{\tt DOI:}\penalty0{#1}\ }
  \fi
\ifx \showISBNx    \undefined \def \showISBNx     #1{\unskip}     \fi
\ifx \showISBNxiii \undefined \def \showISBNxiii  #1{\unskip}     \fi
\ifx \showISSN     \undefined \def \showISSN      #1{\unskip}     \fi
\ifx \showLCCN     \undefined \def \showLCCN      #1{\unskip}     \fi
\ifx \shownote     \undefined \def \shownote      #1{#1}          \fi
\ifx \showarticletitle \undefined \def \showarticletitle #1{#1}   \fi
\ifx \showURL      \undefined \def \showURL       #1{#1}          \fi

\bibitem{Barr:1984}
{Alan~H. Barr}. 1984.
\newblock \showarticletitle{Global and Local Deformations of Solid Primitives}.
\newblock {\em SIGGRAPH Computer Graphics\/} {18}, 3 (Jan. 1984), 21--30.
\newblock
\showISSN{0097-8930}
\showDOI{%
\url{http://dx.doi.org/10.1145/964965.808573}}


\bibitem{Bock:2018}
{A. Bock}, {H. Doraiswamy}, {A. Summers}, {and} {C. Silva}. 2018.
\newblock \showarticletitle{TopoAngler: Interactive Topology-Based Extraction
  of Fishes}.
\newblock {\em IEEE Transactions on Visualization and Computer Graphics\/}
  {24}, 1 (2018), 812--821.
\newblock
\showISSN{1077-2626}
\showDOI{%
\url{http://dx.doi.org/10.1109/TVCG.2017.2743980}}


\bibitem{PMP:2010}
{Mario Botsch}, {Leif Kobbelt}, {Mark Pauly}, {Pierre Alliez}, {and} {Bruno
  Levy}. 2010.
\newblock {\em Polygon Mesh Processing}.
\newblock AK Peters.
\newblock
\showISBNx{978-1-56881-426-1}


\bibitem{Campen:2016}
{Marcel Campen}, {Cl\'{a}udio~T. Silva}, {and} {Denis Zorin}. 2016.
\newblock \showarticletitle{Bijective Maps from Simplicial Foliations}.
\newblock {\em ACM Transactions on Graphics\/} {35}, 4, Article 74 (2016), 15
  pages.
\newblock
\showISSN{0730-0301}
\showDOI{%
\url{http://dx.doi.org/10.1145/2897824.2925890}}


\bibitem{CSA03}
{H. Carr}, {J. Snoeyink}, {and} {U. Axen}. 2003.
\newblock \showarticletitle{Computing {C}ontour {T}rees in {A}ll {D}imensions}.
\newblock {\em Comput. Geom. Theory Appl.\/} {24}, 2 (2003), 75--94.
\newblock


\bibitem{Chen:2003}
{M. Chen}, {D. Silver}, {A.~S. Winter}, {V. Singh}, {and} {N. Cornea}. 2003.
\newblock \showarticletitle{Spatial Transfer Functions: A Unified Approach to
  Specifying Deformation in Volume Modeling and Animation}. In {\em Proceedings
  of the Workshop on Volume Graphics}. 35--44.
\newblock
\showISBNx{1-58113-745-1}
\showDOI{%
\url{http://dx.doi.org/10.1145/827051.827056}}


\bibitem{Chen:2018}
{Zhen Chen}, {Daniele Panozzo}, {and} {J{\'{e}}r{\'{e}}mie Dumas}. 2018.
\newblock \showarticletitle{Half-Space Power Diagrams and Discrete Surface
  Offsets}.
\newblock {\em CoRR\/}  {abs/1804.08968} (2018).
\newblock
\showURL{%
\url{http://arxiv.org/abs/1804.08968}}


\bibitem{Choi2019}
{Minsuk Choi}, {Cheonbok Park}, {Soyoung Yang}, {Yonggyu Kim}, {Jaegul Choo},
  {and} {Sungsoo~Ray Hong}. 2019.
\newblock \showarticletitle{AILA: Attentive Interactive Labeling Assistant for
  Document Classification Through Attention-Based Deep Neural Networks}. In
  {\em Proceedings of the 2019 CHI Conference on Human Factors in Computing
  Systems} {\em (CHI '19)}. ACM, New York, NY, USA, Article 230, 12 pages.
\newblock
\showISBNx{978-1-4503-5970-2}
\showDOI{%
\url{http://dx.doi.org/10.1145/3290605.3300460}}


\bibitem{Claici:2017}
{S. Claici}, {M. Bessmeltsev}, {S. Schaefer}, {and} {J. Solomon}. 2017.
\newblock \showarticletitle{Isometry‐Aware Preconditioning for Mesh
  Parameterization}.
\newblock {\em Computer Graphics Forum\/} {36}, 5 (2017), 37--47.
\newblock
\showDOI{%
\url{http://dx.doi.org/10.1111/cgf.13243}}


\bibitem{Coquillart:1990}
{Sabine Coquillart}. 1990.
\newblock \showarticletitle{Extended Free-form Deformation: A Sculpturing Tool
  for 3D Geometric Modeling}.
\newblock {\em SIGGRAPH Computer Graphics\/} {24}, 4 (1990), 187--196.
\newblock
\showISSN{0097-8930}
\showDOI{%
\url{http://dx.doi.org/10.1145/97880.97900}}


\bibitem{correa2007illustrative}
{C. {Correa}}, {D. {Silver}}, {and} {M. {Chen}}. 2007.
\newblock \showarticletitle{Illustrative Deformation for Data Exploration}.
\newblock {\em IEEE Transactions on Visualization and Computer Graphics\/}
  {13}, 6 (2007), 1320--1327.
\newblock
\showISSN{1077-2626}
\showDOI{%
\url{http://dx.doi.org/10.1109/TVCG.2007.70565}}


\bibitem{correa2007volume}
{Carlos~D Correa}, {Deborah Silver}, {and} {Min Chen}. 2007.
\newblock \showarticletitle{Volume Deformation via Scattered Data
  Interpolation.}. In {\em Volume Graphics}. 91--98.
\newblock


\bibitem{scan-all-fish-science}
{Ryan Cross}.
\newblock New 3D scanning campaign will reveal 20,000 animals in stunning
  detail.
\newblock   (????).
\newblock


\bibitem{Dean2012}
{Mason Dean}, {Dan Huber}, {Brian Goo}, {Nicole Danos}, {Kenshu Shimada}, {and}
  {Adam Summers}. 2012.
\newblock \showarticletitle{On the jaws of lamniform sharks}.
\newblock {\em Integrative and Comparative Biology\/}  {52} (2012).
\newblock


\bibitem{Degener:2003}
{P. Degener}, {J. Meseth}, {and} {R. Klein}. 2003.
\newblock \showarticletitle{An Adaptable Surface Parameterization Method}. In
  {\em Proceedings of the 12th International Meshing Roundtable}. 201--213.
\newblock


\bibitem{Fu:2016}
{Xiao-Ming Fu} {and} {Yang Liu}. 2016.
\newblock \showarticletitle{Computing Inversion-free Mappings by Simplex
  Assembly}.
\newblock {\em ACM Transactions on Graphics\/} {35}, 6, Article 216 (2016), 12
  pages.
\newblock
\showISSN{0730-0301}
\showDOI{%
\url{http://dx.doi.org/10.1145/2980179.2980231}}


\bibitem{Fu:2015}
{Xiao-Ming Fu}, {Yang Liu}, {and} {Baining Guo}. 2015.
\newblock \showarticletitle{Computing Locally Injective Mappings by Advanced
  MIPS}.
\newblock {\em ACM Transactions on Graphics\/} {34}, 4, Article 71 (2015), 12
  pages.
\newblock


\bibitem{gao2016structured}
{Xifeng Gao}, {Tobias Martin}, {Sai Deng}, {Elaine Cohen}, {Zhigang Deng},
  {and} {Guoning Chen}. 2016.
\newblock \showarticletitle{Structured volume decomposition via generalized
  sweeping}.
\newblock {\em IEEE Transactions on Visualization and Computer Graphics\/}
  {22}, 7 (2016), 1899--1911.
\newblock


\bibitem{Hall2018}
{K.~C. Hall}, {P~J. Hundt}, {J.~D. Swenson}, {A.~P. Summers}, {and} {K.~D.
  Crow}. 2018.
\newblock \showarticletitle{The evolution of underwater flight: The
  redistribution of pectoral fin rays, in manta rays and their relatives
  (Myliobatidae)}.
\newblock {\em Journal of Morphology\/} {270}, 8 (2018), 1155--1170.
\newblock
\showDOI{%
\url{http://dx.doi.org/https://doi.org/10.1002/jmor.20837}}


\bibitem{Hong2014}
{Sungsoo~(Ray) Hong}, {Yea-Seul Kim}, {Jong-Chul Yoon}, {and} {Cecilia~R.
  Aragon}. 2014.
\newblock \showarticletitle{Traffigram: Distortion for Clarification via
  Isochronal Cartography}. In {\em Proceedings of the SIGCHI Conference on
  Human Factors in Computing Systems} {\em (CHI '14)}. ACM, New York, NY, USA,
  907--916.
\newblock
\showISBNx{978-1-4503-2473-1}
\showDOI{%
\url{http://dx.doi.org/10.1145/2556288.2557224}}


\bibitem{Hormann:2000}
{K. Hormann} {and} {G. Greiner}. 2000.
\newblock \showarticletitle{{MIPS}: An Efficient Global Parametrization
  Method}.
\newblock In {\em Curve and Surface Design}. 153--162.
\newblock


\bibitem{skinningcourse:2014}
{Alec Jacobson}, {Zhigang Deng}, {Ladislav Kavan}, {and} {JP Lewis}. 2014.
\newblock \showarticletitle{Skinning: Real-time Shape Deformation}. In {\em ACM
  SIGGRAPH 2014 Courses}.
\newblock


\bibitem{Johnson:2004:VH:993936}
{Christopher Johnson} {and} {Charles Hansen}. 2004.
\newblock {\em Visualization Handbook}.
\newblock
\showISBNx{012387582X}


\bibitem{Kolmann2018}
{M.~A. Kolmann}, {J.~M. Huie}, {K. Evans}, {and} {A.~P. Summers}. 2018.
\newblock \showarticletitle{Specialized specialists and the narrow niche
  fallacy: a tale of scale-feeding fishes}.
\newblock {\em Royal Society Open Science\/} (2018).
\newblock
\showDOI{%
\url{http://dx.doi.org/10.1098/rsos.171581}}


\bibitem{Kovalsky:2015}
{Shahar~Z. Kovalsky}, {Noam Aigerman}, {Ronen Basri}, {and} {Yaron Lipman}.
  2015.
\newblock \showarticletitle{Large-scale Bounded Distortion Mappings}.
\newblock {\em ACM Trans. Graph.\/} {34}, 6, Article 191 (Oct. 2015), 10 pages.
\newblock
\showISSN{0730-0301}
\showDOI{%
\url{http://dx.doi.org/10.1145/2816795.2818098}}


\bibitem{Kovalsky:2016}
{Shahar~Z. Kovalsky}, {Meirav Galun}, {and} {Yaron Lipman}. 2016.
\newblock \showarticletitle{Accelerated Quadratic Proxy for Geometric
  Optimization}.
\newblock {\em ACM Transactions on Graphics\/} {35}, 4, Article 134 (2016), 11
  pages.
\newblock
\showISSN{0730-0301}
\showDOI{%
\url{http://dx.doi.org/10.1145/2897824.2925920}}


\bibitem{Kwon:2018}
{Soonhyeon Kwon}, {Younguk Kim}, {Kihyuk Kim}, {and} {Sungkil Lee}. 2018.
\newblock \showarticletitle{Heterogeneous volume deformation and animation
  authoring with density-aware moving least squares}.
\newblock {\em Computer Animation and Virtual Worlds\/} {29}, 1 (2018).
\newblock
\showURL{%
\url{https://onlinelibrary.wiley.com/doi/abs/10.1002/cav.1784}}


\bibitem{labelle2007isosurface}
{Fran{\c{c}}ois Labelle} {and} {Jonathan~Richard Shewchuk}. 2007.
\newblock \showarticletitle{Isosurface stuffing: fast tetrahedral meshes with
  good dihedral angles}. In {\em ACM Transactions on Graphics}, Vol.~26. 57.
\newblock


\bibitem{Li:2007}
{Wilmot Li}, {Lincoln Ritter}, {Maneesh Agrawala}, {Brian Curless}, {and}
  {David Salesin}. 2007b.
\newblock \showarticletitle{Interactive Cutaway Illustrations of Complex 3D
  Models}.
\newblock {\em ACM Transactions on Graphics\/} {26}, 3, Article 31 (July 2007).
\newblock
\showISSN{0730-0301}
\showDOI{%
\url{http://dx.doi.org/10.1145/1276377.1276416}}


\bibitem{Li:2007:Harmonic}
{Xin Li}, {Xiaohu Guo}, {Hongyu Wang}, {Ying He}, {Xianfeng Gu}, {and} {Hong
  Qin}. 2007a.
\newblock \showarticletitle{Harmonic Volumetric Mapping for Solid Modeling
  Applications}. In {\em Proceedings of the ACM Symposium on Solid and Physical
  Modeling}. 109--120.
\newblock
\showISBNx{978-1-59593-666-0}
\showDOI{%
\url{http://dx.doi.org/10.1145/1236246.1236263}}


\bibitem{Lipman:2012}
{Yaron Lipman}. 2012.
\newblock \showarticletitle{Bounded Distortion Mapping Spaces for Triangular
  Meshes}.
\newblock {\em ACM Transactions on Graphics\/} {31}, 4 (2012), 108:1--108:13.
\newblock


\bibitem{liu2014survey}
{Shixia Liu}, {Weiwei Cui}, {Yingcai Wu}, {and} {Mengchen Liu}. 2014.
\newblock \showarticletitle{A survey on information visualization: recent
  advances and challenges}.
\newblock {\em The Visual Computer\/} {30}, 12 (2014), 1373--1393.
\newblock


\bibitem{Livesu:2017}
{Marco Livesu}, {Marco Attene}, {Giuseppe Patan\'{e}}, {and} {Michela
  Spagnuolo}. 2017.
\newblock \showarticletitle{Explicit cylindrical maps for general tubular
  shapes}.
\newblock {\em Computer-Aided Design\/}  {90} (2017), 27--36.
\newblock
\showISSN{0010-4485}
\showDOI{%
\url{http://dx.doi.org/https://doi.org/10.1016/j.cad.2017.05.002}}


\bibitem{Meier:2005}
{U. Meier}, {O. L\'{o}pez}, {C. Monserrat}, {M.~C. Juan}, {and} {M.
  Alca\~{n}iz}. 2005.
\newblock \showarticletitle{Real-time Deformable Models for Surgery Simulation:
  A Survey}.
\newblock {\em Computer Methods and Programs in Biomedicine\/} {77}, 3 (2005),
  183--197.
\newblock
\showISSN{0169-2607}
\showDOI{%
\url{http://dx.doi.org/10.1016/j.cmpb.2004.11.002}}


\bibitem{MRH08}
{J{\"o}rg Mensmann}, {Timo Ropinski}, {and} {Klaus Hinrichs}. 2008.
\newblock \showarticletitle{{Interactive Cutting Operations for Generating
  Anatomical Illustrations from Volumetric Data Sets}}.
\newblock {\em Journal of WSCG\/} {16}, 2 (2008), 89--96.
\newblock


\bibitem{Nakao:2010}
{M. {Nakao}}, {K.~W.~C. {Hung}}, {S. {Yano}}, {K. {Yoshimura}}, {and} {K.
  {Minato}}. 2010.
\newblock \showarticletitle{Adaptive proxy geometry for direct volume
  manipulation}. In {\em 2010 IEEE Pacific Visualization Symposium}. 161--168.
\newblock
\showISSN{2165-8765}
\showDOI{%
\url{http://dx.doi.org/10.1109/PACIFICVIS.2010.5429597}}


\bibitem{Nakao:2014}
{Megumi Nakao}, {Yuya Oda}, {Kojiro Taura}, {and} {Kotaro Minato}. 2014.
\newblock \showarticletitle{Direct volume manipulation for visualizing
  intraoperative liver resection process}.
\newblock {\em Computer Methods and Programs in Biomedicine\/} {113}, 3 (2014),
  725--735.
\newblock
\showISSN{0169-2607}
\showDOI{%
\url{http://dx.doi.org/https://doi.org/10.1016/j.cmpb.2013.12.004}}


\bibitem{Nealen:2006}
{Andrew Nealen}, {Matthias Mueller}, {Richard Keiser}, {Eddy Boxerman}, {and}
  {Mark Carlson}. 2006.
\newblock \showarticletitle{{Physically Based Deformable Models in Computer
  Graphics}}.
\newblock {\em Computer Graphics Forum\/} (2006).
\newblock
\showDOI{%
\url{http://dx.doi.org/10.1111/j.1467-8659.2006.01000.x}}


\bibitem{Ono2019}
{Jorge Piazentin~Ono}, {Arvi Gjoka}, {Justin Salamon}, {Carlos Dietrich}, {and}
  {Claudio~T. Silva}. 2019.
\newblock \showarticletitle{HistoryTracker: Minimizing Human Interactions in
  Baseball Game Annotation}. In {\em Proceedings of the 2019 CHI Conference on
  Human Factors in Computing Systems} {\em (CHI '19)}. ACM, New York, NY, USA,
  Article 63, 12 pages.
\newblock
\showISBNx{978-1-4503-5970-2}
\showDOI{%
\url{http://dx.doi.org/10.1145/3290605.3300293}}


\bibitem{Poranne:2017}
{Roi Poranne}, {Marco Tarini}, {Sandro Huber}, {Daniele Panozzo}, {and} {Olga
  Sorkine-Hornung}. 2017.
\newblock \showarticletitle{Autocuts: Simultaneous Distortion and Cut
  Optimization for UV Mapping}.
\newblock {\em ACM Transactions on Graphics\/} {36}, 6, Article 215 (2017), 11
  pages.
\newblock
\showISSN{0730-0301}
\showDOI{%
\url{http://dx.doi.org/10.1145/3130800.3130845}}


\bibitem{Rabinovich:2017}
{Michael Rabinovich}, {Roi Poranne}, {Daniele Panozzo}, {and} {Olga
  Sorkine-Hornung}. 2017.
\newblock \showarticletitle{Scalable Locally Injective Mappings}.
\newblock {\em ACM Transactions on Graphics\/} {36}, 2, Article 16 (2017), 16
  pages.
\newblock
\showISSN{0730-0301}
\showDOI{%
\url{http://dx.doi.org/10.1145/2983621}}


\bibitem{Schneider:2015}
{Teseo Schneider} {and} {Kai Hormann}. 2015.
\newblock \showarticletitle{Smooth bijective maps between arbitrary planar
  polygons}.
\newblock {\em Computer Aided Geometric Design\/} {35--36}, C (2015), 243--354.
\newblock
\newblock
\shownote{Proceedings of GMP.}


\bibitem{Schuller:2013}
{Christian Sch\"{u}ller}, {Ladislav Kavan}, {Daniele Panozzo}, {and} {Olga
  Sorkine-Hornung}. 2013.
\newblock \showarticletitle{Locally Injective Mappings}. In {\em Proceedings of
  the Symposium on Geometry Processing}. 125--135.
\newblock
\showDOI{%
\url{http://dx.doi.org/10.1111/cgf.12179}}


\bibitem{Sederberg:1986}
{Thomas~W. Sederberg} {and} {Scott~R. Parry}. 1986.
\newblock \showarticletitle{Free-form Deformation of Solid Geometric Models}.
\newblock {\em SIGGRAPH Computer Graphics\/} {20}, 4 (1986), 151--160.
\newblock
\showISSN{0097-8930}
\showDOI{%
\url{http://dx.doi.org/10.1145/15886.15903}}


\bibitem{Shtengel:2017}
{Anna Shtengel}, {Roi Poranne}, {Olga Sorkine-Hornung}, {Shahar~Z. Kovalsky},
  {and} {Yaron Lipman}. 2017.
\newblock \showarticletitle{Geometric Optimization via Composite Majorization}.
\newblock {\em ACM Transactions on Graphics\/} {36}, 4, Article 38 (2017), 11
  pages.
\newblock
\showDOI{%
\url{http://dx.doi.org/10.1145/3072959.3073618}}


\bibitem{Sifakis:2012}
{Eftychios Sifakis} {and} {Jernej Barbic}. 2012.
\newblock \showarticletitle{{FEM} Simulation of {3D} Deformable Solids: A
  Practitioner's Guide to Theory, Discretization and Model Reduction}. In {\em
  ACM SIGGRAPH 2012 Courses}. Article 20, 50 pages.
\newblock
\showISBNx{978-1-4503-1678-1}
\showURL{%
\url{http://doi.acm.org/10.1145/2343483.2343501}}


\bibitem{Smith:2015}
{Jason Smith} {and} {Scott Schaefer}. 2015.
\newblock \showarticletitle{Bijective Parameterization with Free Boundaries}.
\newblock {\em ACM Transactions on Graphics\/} {34}, 4, Article 70 (2015), 9
  pages.
\newblock
\showISSN{0730-0301}


\bibitem{ARAP_modeling:2007}
{Olga Sorkine} {and} {Marc Alexa}. 2007.
\newblock \showarticletitle{As-Rigid-As-Possible Surface Modeling}. In {\em
  Proceedings of EUROGRAPHICS/ACM SIGGRAPH Symposium on Geometry Processing}.
  109--116.
\newblock


\bibitem{Stocker2019}
{M.~R. Stocker}, {S.~J. Nesbitt}, {B.~T. Kligman}, {D.~J. Paluh}, {A.~D.
  Marsh}, {D.~C. Blackburn}, {and} {Parker~W. G.} 2019.
\newblock \showarticletitle{The earliest equatorial record of frogs from the
  Late Triassic of Arizona}.
\newblock {\em Biology Letters\/} (2019).
\newblock


\bibitem{sun2013survey}
{Guo-Dao Sun}, {Ying-Cai Wu}, {Rong-Hua Liang}, {and} {Shi-Xia Liu}. 2013.
\newblock \showarticletitle{A survey of visual analytics techniques and
  applications: State-of-the-art research and future challenges}.
\newblock {\em Journal of Computer Science and Technology\/} {28}, 5 (2013),
  852--867.
\newblock


\bibitem{Tagliasacchi:2016}
{Andrea Tagliasacchi}, {Thomas Delame}, {Michela Spagnuolo}, {Nina Amenta},
  {and} {Alexandru Telea}. 2016.
\newblock \showarticletitle{{3D Skeletons: A State-of-the-Art Report}}.
\newblock {\em {Computer Graphics Forum}\/} {35}, 2 (2016), 573--597.
\newblock
\showDOI{%
\url{http://dx.doi.org/10.1111/cgf.12865}}


\bibitem{Volume_Wires}
{Simon~J. Walton} {and} {Mark~W. Jones}. 2006.
\newblock \showarticletitle{Volume Wires: A Framework for Empirical Non-linear
  Deformation of Volumetric Datasets}.
\newblock {\em Journal of WSCG\/} {14}, 1--3 (2006), 81--88.
\newblock
\showISSN{1213-6972}


\bibitem{Wang:2004:harmonic}
{Yalin Wang}, {Xianfeng Gu}, {and} {Shing-Tung Yau}. 2003.
\newblock \showarticletitle{Volumetric harmonic map}.
\newblock {\em Communications in Information \& Systems\/} {3}, 3 (2003),
  191--202.
\newblock


\bibitem{Watkins2018}
{Gregory Watkins-Colwell}, {Kevin Love}, {Zachary Randall}, {Doug Boyer},
  {Julie Winchester}, {Edward Stanley}, {and} {David Blackburn}. 2018.
\newblock \showarticletitle{The Walking Dead: Status Report, Data Workflow and
  Best Practices of the oVert Thematic Collections Network}.
\newblock {\em Biodiversity Information Science and Standards\/}  {2} (2018).
\newblock
\showDOI{%
\url{http://dx.doi.org/10.3897/biss.2.26078}}


\bibitem{Witkins:1997}
{Andrew Witkin}. 1997.
\newblock \showarticletitle{Physically Based Modeling: Principles and Practice
  Particle System Dynamics}.
\newblock {\em ACM Siggraph Course\/} (1997).
\newblock


\end{thebibliography}

\end{document}


\title{Supplementary Material for \\Unwind: Interactive Fish Straightening}

\author{Francis Williams, Alexander Bock, Harish Doraiswamy, Cassandra Donatelli, Kayla Hall, Adam Summers, Daniele Panozzo, Cl\'audio T. Silva
}

\maketitle

\subsection{Extra Results}\label{sec:extra_results}
Below are more results of fishes straightened using unwind by a marine-biologist expert user.

\begin{figure}[b]
    \centering
    Input
    \includegraphics[width=\columnwidth]{figures/results/AmmodytesPersonatusTwisted001/input_side-fs8.png}
    \includegraphics[width=\columnwidth]{figures/results/AmmodytesPersonatusTwisted001/input_top-fs8.png}
    Output
    \includegraphics[width=\columnwidth]{figures/results/AmmodytesPersonatusTwisted001/output_side-fs8.png}
    \includegraphics[width=\columnwidth]{figures/results/AmmodytesPersonatusTwisted001/output_top-fs8.png}
    \caption{Input and output of our pipeline. The top image shows shows a side view and the bottom shows a top view.}
    \label{fig:fish01}
\end{figure}
\begin{figure}
    \centering
    Input
    \includegraphics[width=\columnwidth]{figures/results/LumpenusSagittaTwisted002/input_side-fs8.png}
    \includegraphics[width=\columnwidth]{figures/results/LumpenusSagittaTwisted002/input_top-fs8.png}
    Output
    \includegraphics[width=\columnwidth]{figures/results/LumpenusSagittaTwisted002/output_side-fs8.png}
    \includegraphics[width=\columnwidth]{figures/results/LumpenusSagittaTwisted002/output_top-fs8.png}
    \caption{Input and output of our pipeline. The top image shows shows a side view and the bottom shows a top view.}
    \label{fig:fish08}
\end{figure}
\begin{figure}
    \centering
    Input
    \includegraphics[width=\columnwidth]{figures/results/PholisLaetaTwisted001/input_side-fs8.png}
    \includegraphics[width=\columnwidth]{figures/results/PholisLaetaTwisted001/input_top-fs8.png}
    Output
    \includegraphics[width=\columnwidth]{figures/results/PholisLaetaTwisted001/output_side-fs8.png}
    \includegraphics[width=\columnwidth]{figures/results/PholisLaetaTwisted001/output_top-fs8.png}
    \caption{Input and output of our pipeline. The top image shows shows a side view and the bottom shows a top view.}
    \label{fig:fish09}
\end{figure}
\begin{figure}
    \centering
    Input
    \includegraphics[width=\columnwidth]{figures/results/LumpenusSagittaTwisted001/input_side-fs8.png}
    \includegraphics[width=\columnwidth]{figures/results/LumpenusSagittaTwisted001/input_top-fs8.png}
    Output
    \includegraphics[width=\columnwidth]{figures/results/LumpenusSagittaTwisted001/output_side-fs8.png}
    \includegraphics[width=\columnwidth]{figures/results/LumpenusSagittaTwisted001/output_top-fs8.png}
    \caption{Input and output of our pipeline. The top image shows shows a side view and the bottom shows a top view.}
    \label{fig:fish07}
\end{figure}
\begin{figure}
    \centering
    Input
    \includegraphics[width=0.95\columnwidth]{figures/results/AnoplarchusPurpurescensTwisted001/input_side-fs8.png}
    \includegraphics[width=0.95\columnwidth]{figures/results/AnoplarchusPurpurescensTwisted001/input_top-fs8.png}
    Output
    \includegraphics[width=0.95\columnwidth]{figures/results/AnoplarchusPurpurescensTwisted001/output_side-fs8.png}
    \includegraphics[width=0.95\columnwidth]{figures/results/AnoplarchusPurpurescensTwisted001/output_top-fs8.png}
    \caption{Input and output of our pipeline. The top image shows shows a side view and the bottom shows a top view.}
    \label{fig:fish03}
\end{figure}
\begin{figure}
    \centering
    Input
    \includegraphics[width=\columnwidth]{figures/results/XiphisterMucosusTwisted001/input_side-fs8.png}
    \includegraphics[width=\columnwidth]{figures/results/XiphisterMucosusTwisted001/input_top-fs8.png}
    Output
    \includegraphics[width=\columnwidth]{figures/results/XiphisterMucosusTwisted001/output_side-fs8.png}
    \includegraphics[width=\columnwidth]{figures/results/XiphisterMucosusTwisted001/output_top-fs8.png}
    \caption{Input and output of our pipeline. The top image shows shows a side view and the bottom shows a top view.}
    \label{fig:fish11}
\end{figure}

\begin{figure}
    \centering
    Input\\
    \includegraphics[width=0.5\columnwidth]{figures/results/AnoplarchusPurpurescensBent001/input_side-fs8.png}\\
    \includegraphics[width=0.5\columnwidth]{figures/results/AnoplarchusPurpurescensBent001/input_top-fs8.png}\\
    Output\\
    \includegraphics[width=0.5\columnwidth]{figures/results/AnoplarchusPurpurescensBent001/output_side-fs8.png}\\
    \includegraphics[width=0.5\columnwidth]{figures/results/AnoplarchusPurpurescensBent001/output_top-fs8.png}
    \caption{Despite missing parts of the scan, our system can produce a straightened exemplar.}
    \label{fig:fish02}
\end{figure}
\begin{figure}
    \centering
    Input\\
    \includegraphics[width=0.5\columnwidth]{figures/results/ApodichthysFlavidusBent001/input_side-fs8.png}\\
    \includegraphics[width=0.5\columnwidth]{figures/results/ApodichthysFlavidusBent001/input_top-fs8.png}\\
    Output\\
    \includegraphics[width=0.5\columnwidth]{figures/results/ApodichthysFlavidusBent001/output_side-fs8.png}\\
    \includegraphics[width=0.5\columnwidth]{figures/results/ApodichthysFlavidusBent001/output_top-fs8.png}\\
    \caption{This fish was purposefully bent to an extreme position to evaluate the capabilities of our system in the presence of large deformations .}
    \label{fig:fish04}
\end{figure}

\subsection{Transcript of Interview With Expert User}\label{sec:expert_interview}
What follows is the transcript of an interview with one of our expert users in the Tytell Lab at Tufts University.\\

\noindent
\textbf{Q:} \textit{How long did it take you and your student to get familiar with the workflow? }\\
\textbf{A:} We chatted about it and I showed her some stuff for maybe 5-8 minutes. It was pretty quick!
\\\\
\textbf{Q:} \textit{What were any difficulties that you and your student faced to begin with, is that now resolved or do you still have any issues?}\\
\textbf{A:} The only issue she ran into was with saving. Since it just kind of closes, she thought it didn't save and started to re-do it. I had to explain that closing wasn't crashing, that's what it does when saving. For me, the only issue I was having was with missing parts of fish, but I think that's been solved.
\\\\
\textbf{Q:} \textit{What do you guys like about this tool?}\\
\textbf{A:} I like that it's easy to use and a lot more intuitive than the segmenting tool (which I also liked but I was one of the only people who could figure out how to use it).
\\\\
\textbf{Q:} \textit{How does having this tool make your lives easy?}\\
\textbf{A:} The tool is really useful for doing morphometrics analysis, especially when we are trying to look at things in a specific anatomical plane. Before, we had to have versions of the scan rotated at different angles to measure things in a somewhat straight line. Now we can just straighten a single scan which saves a lot of time, storage space, and confusion with different versions of files. On a less scientific note, it makes creating figures a lot easier. We don't have to spend hours looking for the perfect scan, we can just fix any one we have. 
\\\\
\textbf{Q:} \textit{How frequently do you plan to use this tool?}\\
\textbf{A:} I would use it every time I go to analyze CT scans of my fish. My species especially are really bendy and flexible, so it's not hard to accidentally bend or twist parts while preparing a scan. So for me I'll use it all the time.
\\\\
\textbf{Q:} \textit{Suggestions on what to improve?}\\
\textbf{A:} I think the biggest thing would be the option for different import and export formats. Lots of people end up with tiff stacks, so being able to import them would be awesome. For export, being able to save a DICOM stack or a NRRD files would be useful. 
\\\\
\textbf{Q:} \textit{Any other thoughts / feedback etc?}\\
\textbf{A:} I was talking to my adviser and he was saying another thing folks might actually find useful is the ability to actually bend fish to specific angles. For example, my adviser has scans of fish with muscle fibers stained. It might be interesting to take a straight fish, and then see what the difference in muscle fiber length would look like in a bent fish. Just as an interesting "future directions" thing.